\documentclass[11pt,a4paper]{article}
\usepackage{jcappub}
\usepackage{subcaption}
\usepackage{mathtools}

\author[a]{Chad Briddon,}
\author[a]{Clare Burrage,}
\author[a]{Adam Moss,}
\author[a]{and Andrius Tamosiunas}

\affiliation[a]{School of Physics and Astronomy, University of Nottingham, Nottingham, NG7 2RD,\\ United Kingdom}

\emailAdd{chad.briddon@nottingham.ac.uk}
\emailAdd{clare.burrage@nottingham.ac.uk}
\emailAdd{adam.moss@nottingham.ac.uk}
\emailAdd{andrius.tamosiunas@nottingham.ac.uk}

\title{SELCIE: A tool for investigating the chameleon field of arbitrary sources}

\abstract{The chameleon model is a modified gravity theory that introduces an additional scalar field that couples to matter through a conformal coupling. This `chameleon field' possesses a screening mechanism through a nonlinear self-interaction term which allows the field to affect cosmological observables in diffuse environments whilst still being consistent with current local experimental constraints. Due to the self-interaction term the equations of motion of the field are nonlinear and therefore difficult to solve analytically. The analytic solutions that do exist in the literature are either approximate solutions and or only apply to highly symmetric systems. In this work we introduce the software package SELCIE (\url{https://github.com/C-Briddon/SELCIE.git}). This package equips the user with tools to construct an arbitrary system of mass distributions and then to calculate the corresponding solution to the chameleon field equation. It accomplishes this by using the finite element method and either the Picard or Newton nonlinear solving methods. We compared the results produced by SELCIE with analytic results from the literature including discrete and continuous density distributions. We found strong (sub-percentage) agreement between the solutions calculated by SELCIE and the analytic solutions.}

\begin{document}
\maketitle
\flushbottom
\section{Introduction}
\label{sec:Introduction}

With the discovery of the Higgs boson in 2012 \cite{Aad_2012}, we have observed a true scalar field in nature. However, it remains possible that the Higgs is not the only scalar field, and there are observed phenomenon that could be explained by the existence of new scalar fields. For example, from observations of type Ia supernova, baryon acoustic oscillations (BAO), and gravitational waves \cite{Riess:1998cb, Eisenstein_2005, Holz_2005, Abbott_2021}, it has been discovered that the universe is currently in a period of accelerated expansion. The source of this expansion is referred to as dark energy and is currently a mystery of cosmology. One proposed explanation is the introduction of scalar fields, either directly or through modifications to gravity, to explain this behaviour \cite{Slosar:2019flp,Joyce:2014kja, Zlatev:1998tr,Copeland:2006wr}. The confirmation of the existence, or lack thereof, of new scalar fields is therefore vital to our collective understanding of the universe. 

If such a scalar field exists and couples to matter, it can mediate a so called `fifth force' giving rise to many opportunities to search for this new physics \cite{Khoury:2003rn}. At the time of writing, terrestrial and solar system experiments have not detected any fifth forces, and as a result many models that rely on such scalar fields are strongly constrained to have very weak couplings to matter \cite{Wagner:2012ui,Adelberger:2003zx}. An alternative to this fine tuning arises in screened models where nonlinearites can suppress the fifth force dynamically in and around higher density environments \cite{Khoury:2003rn,Joyce:2014kja}. A screened field which is not detected in terrestrial and solar system experiments can still affect large scale cosmological evolution. In recent years much progress has been made in tailoring experiments and observations to search for such theories, and some theories with screening are now also extremely constrained \cite{Burrage:2017qrf,Noller:2020afd}. As a result it is not enough for a theory to possess a screening mechanism for it to be able to reproduce the observed cosmic acceleration and evade local experiments, see for example Ref.~\cite{Wang_2012}. The nonlinearities of the theory mean that properties of the scalar fluctuations (which mediate the fifth force) become background dependent. Assuming that the scalar couples universally to matter, this means that, in non-relativistic environments, properties of the scalar field depend on the local energy density. It is possible to classify screening models by which property of the fundamental scalar becomes density dependent. In the chameleon model \cite{Khoury:2003rn} the scalar field has a mass that increases with the local matter density, while the symmetron model \cite{Hinterbichler:2010es} has a matter coupling that decreases with the local matter density. A third type of screening, known as Vainshtein screening, occurs when the kinectic term in the Lagrangian describing the scalar fluctuations becomes dependent on the environment \cite{Vainshtein:1972sx,Nicolis:2008in,Babichev:2009ee,Deffayet:2011gz}. 

To understand the constraints on screened scalar field theories and identify the parameter space that remains viable, exact solutions to the equations of motion are needed. However, one consequence of the reliance of screened models on nonlinearities is that analytic solutions to the equations of motion become much harder to find. This is especially true for situations where the matter sourcing the field is an irregular shape. For the chameleon model, for example, various approximate analytic solutions exist but only for highly symmetric source shapes such as spheres, planes and ellipsoids \cite{Khoury:2003rn, Burrage_2015, Mota:2006fz}. Even for this small subset of possible source shapes we see that the strength of the fifth force can vary depending on the matter configuration. To compare screened scalar theories with experimental and observational tests, and to optimise these searches, we need the ability to determine the scalar field profiles and corresponding fifth forces for arbitrary sources. 

SELCIE (Screening Equations Linearly Constructed and Iteratively Evaluated) is a software package that provides the user with tools to investigate nonlinear scalar field models such as the symmetron and chameleon in user defined systems such as an irregularly shaped source inside a vacuum chamber, galaxy clusters, etc. To accomplish this SELCIE uses a nonlinear solving method (either the Picard or Newton method \cite{kelley1995iterative}) with the finite element method (FEM) via the software package FEniCS \cite{AlnaesBlechta2015a, LoggMardalEtAl2012a, LoggWells2010a, LoggWellsEtAl2012a}. Through the FEM the field equations can be solved over irregularly spaced meshes. This allows the use of meshes that are more refined in regions where the field variation is largest, allowing us to solve the equations to a greater degree of accuracy with less computing time. Tools to easily construct and optimise these meshes are also provided in SELCIE, using the mesh generating software GMSH \cite{gmsh, Geuzaine2009Gmsh}. From the calculated field solutions, SELCIE can determine the strength of the corresponding fifth force that would be experienced by test particles.

This is not the first time the finite element method, or meshes tailored to experimental configurations, have been used to determine the behaviour of screened scalar fields. A similar approach to solving screened equations of motion using the finite element method was recently taken in Ref.~\cite{Braden_2021}, and this has been used to investigate the existence of screening in UV complete Galileon models \cite{Burrage:2020bxp}. The behaviour of chameleon and symmetron fields inside an experimental chamber has been studied using finite difference and finite element techniques, in Ref.~\cite{Upadhye:2006vi, Elder:2016yxm} leading to new bounds on the parameters of the theory \cite{Elder:2016yxm, Jaffe:2016fsh}. In Ref.~\cite{Elder:2019yyp} the symmetron equations of motion were solved for the experimental set-ups used to search for Casimir forces. 

Currently SELCIE is configured to find solutions for the chameleon equation of motion; however, in principle the methodology used can be generalized to other scalar fields. We chose the chameleon model as our initial model since it is already heavily constrained at dark energy scales. This means using SELCIE to improve experimental or observational searches could allow us to fully rule out this important part of the chameleon parameter space.

In this paper we will demonstrate the effectiveness of SELCIE in solving the chameleon equation of motion in a variety of different scenarios. We start in Section \ref{sec:The Chameleon Model} by introducing the chameleon theory and its equation of motion, and show how these equations can be rescaled for ease of numerical solving. In Section \ref{sec:Finite Element Method} we introduce the FEM, followed in Section \ref{sec:Nonlinear Solvers} by a description of the nonlinear solving methods that we use. Section \ref{sec:Using SELCIE} describes the application of SELCIE and Section \ref{sec:Comparing Code to Past Results} describes how its results compare to existing results. Finally, we conclude with a summary of the results and will briefly discuss the future of SELCIE.

\textbf{Conventions}: In this paper all calculations will be performed in natural units ($c=1$ and $\hbar=1$). The metric convention in this paper is $\eta_{\mu \nu} = {\rm diag}(-1,1,1,1)$. The reduced Planck mass is defined as $M_{\rm pl}^2 = (8 \pi G_N)^{-1}$, where $G_N$ is Newton's gravitational constant. Partial derivatives $\frac{\partial f}{\partial x}$ and $\frac{\partial^2 f}{\partial x \partial y}$, will be denoted by $f_{,x}$ and $f_{,xy}$ respectively.

\section{The chameleon model}
\label{sec:The Chameleon Model}

The chameleon field, $\phi$, has an action
\begin{equation}
\label{eq:chameleon action}
    S = \int dx^4 \sqrt{-g} \left( \frac{M_{\rm pl}^2}{2} R - \frac{1}{2} \phi^{,\mu} \phi_{,\mu} - V(\phi)\right) - \int dx^4 \mathcal{L}_m(\varphi_m^{(i)}, \Tilde{g}^{(i)}_{\mu \nu}),
\end{equation}
where $V(\phi)$ is the field potential and $R$ is the Ricci scalar \cite{Khoury:2003rn, Pernot-Borras:2020jev}. The matter fields, $\varphi_m^{(i)}$, are described by the Lagrangian $\mathcal{L}_m(\varphi_m^{(i)}, \Tilde{g}^{(i)}_{\mu \nu})$ and the `$i$-th' matter species moves on geodesics of the Jordan frame metric, $\Tilde{g}^{(i)}_{\mu \nu}$. This metric is related to the Einstein frame metric, $g_{\mu \nu}$, (where $g$ is its determinant) by:
\begin{equation}
\label{eq:Jordan to Einstein frame}
    \Tilde{g}^{(i)}_{\mu \nu} = A^2_i(\phi) g_{\mu \nu}.
\end{equation}
Applying the least action principle to equation (\ref{eq:chameleon action}) gives the equation of motion for the field,
\begin{equation}
    \label{eq:eom}
    \Box \phi = V_{,\phi} - \sum_i \frac{\beta_i}{M_{\rm pl}} A_{i}^4 T^{(i)}_{\mu \nu} \Tilde{g}^{\mu \nu}_{(i)},
\end{equation}
where $\Box$ is the d'Alembert operator, the coupling parameter $\beta$ is defined as
\begin{equation}
\label{eq:def coupling constant}
    \beta_i \left(\phi\right) = M_{\rm pl} \frac{d(\ln A_i)}{d \phi},
\end{equation}
and the energy-momentum tensor, $T^{(i)}_{\mu \nu}$, is defined
\begin{equation}
    \label{eq:def T}
    T^{(i)}_{\mu \nu} = \frac{2}{\sqrt{-\Tilde{g}_{(i)}}} \frac{\partial \mathcal{L}_m}{\partial \Tilde{g}^{\mu \nu}_{(i)}}.
\end{equation}
For non-relativistic matter $T^{(i)}_{\mu \nu} \Tilde{g}^{\mu \nu}_{(i)} \approx - \rho_i A_{i}^{-3}$, where $\rho_i$ is the energy density in the Einstein frame. Applying this approximation to equation (\ref{eq:eom}), we find an equation of motion in Klein-Gordan form, $\Box \phi = V_{{\rm eff},\phi}$, where the effective potential is defined as
\begin{equation}
    \label{eq:generic_Veff_form}
        V_{\rm eff}(\phi) = V(\phi) + \sum_i \rho_i A_{i}\left(\phi\right).
\end{equation}
Assuming $\beta_i$ is constant in $\phi$, which results from a conformal coupling of $A_i\left(\phi\right) = e^{\beta_i \phi/M_{\rm pl}}$, if $V(\phi)$ is taken to be monotonically decreasing, equation (\ref{eq:generic_Veff_form}) shows that the effective potential will have a unique minimum for $\beta_i > 0$. This minimum field value, $\phi_{\rm min}$, satisfies
\begin{equation}
    \label{eq:genral phi_min def}
    V_{,\phi}(\phi_{\rm min}) + \sum_i \frac{\beta_i \rho_i}{M_{\rm pl}} e^{\beta_i \phi_{min}/M_{\rm pl}} = 0.
\end{equation}
The effective mass of the field around this minimum is
\begin{equation}
    \label{eq:genral m def}
    m_{\phi}^2 = V_{\rm eff,\phi \phi}(\phi_{\rm min}),
\end{equation}
from which we can find the Compton wavelength of the field; $\lambda_c = m_{\phi}^{-1}$. For an appropriate choice of potential, we see that the mass of the chameleon field is dependent on the value of $\phi_{\rm min}$ which itself depends on $\rho_i$. If the relation between the field mass and the local matter density is such that an increase to the latter results in an increase to the former, then a resulting fifth force will be suppressed in high density regions. As a result, sufficiently large and dense sources exhibit a thin-shell effect where the exterior field is only sourced by matter contained in a thin outer shell \cite{Khoury:2003rn}. Objects such as these are said to be screened and will experience a suppressed fifth force from the field, while unscreened objects will experience the full fifth force. In this way the nonlinearities of the chameleon potential can result in screening of the fifth force. 

In this work we use a chameleon field potential of the form
\begin{equation}
    \label{eq:chameleon potencial}
    V(\phi) = \Lambda^4 \left( 1 + \frac{\Lambda^n}{\phi^n} \right),
\end{equation}
where $\Lambda$ is a mass scale and $n$ is an integer \cite{Khoury:2003rn,Pernot-Borras:2020jev}. We continue to assume that $\beta_{i}$ is independent of $\phi$; however, we will now also assume it to be universal for all matter components and that $\beta \phi \ll M_{\rm pl}$. Using Equation (\ref{eq:genral phi_min def}) we can compute the position in field space of the minimum of the effective potential. For a matter density of $\rho_{0}$, the value of the field that minimises the effective potential is
\begin{equation}
\label{eq:chameleon_min}
    \phi_{0} = \left(\frac{n M_{\rm pl} \Lambda^{n+4}}{\beta \rho_{0}} \right)^{\frac{1}{n+1}}.
\end{equation}
For the purposes of this paper, the chameleon field and matter distribution describing the experimental setup under consideration will be treated as static.\footnote{ In future work we plan to relax this constraint.} The chameleon equation of motion to be solved is therefore:
\begin{equation}
\label{eq:chameleon}
    \nabla^2 \phi = -\frac{n \Lambda^{n+4}}{\phi^{n+1}} + \frac{\beta \rho}{M_{\rm pl}}.
\end{equation}
Once a solution to this equation has been found, the chameleon force experienced by an unscreened test particle of mass $m$ can be computed from the geodesic equation for the metric $\Tilde{g}_{\mu \nu}$ to be
\begin{equation}
\label{eq:force}
    \vec{F}_{\phi} = - \frac{m \beta}{M_{\rm pl}}\vec{\nabla}\phi.
\end{equation}

The equation of motion (\ref{eq:chameleon}) can be rewritten in terms of dimensionless parameters and variables. The field and local density are rescaled using $\phi_{0}$ and $\rho_{0}$, to give the dimensionless quantities $\hat{\rho} = \rho/\rho_{0}$ and $\hat{\phi} = \phi/\phi_{0}$. Spatial distances are rescaled with respect to an arbitrary length scale $L$ so that the Laplacian rescales as $\hat{\nabla}^2 = L^{2}\nabla^2$. Although this length is arbitrary, in practise it is useful to take it to be the characteristic length of the system of interest. The resulting equation of motion is:
\begin{equation}
\label{eq:chameleon_Norm}
    \alpha\hat{\nabla}^2 \hat{\phi} = -\hat{\phi}^{-(n+1)} + \hat{\rho},
\end{equation}
where the dimensionless constant $\alpha$ is defined as
\begin{equation}
\label{eq:alpha_def}
    \alpha \equiv \left( \frac{M_{\rm pl} \Lambda}{L^2 \beta \rho_{0}}\right) \left(  \frac{n M_{\rm pl} \Lambda^3}{\beta \rho_{0}} \right)^{\frac{1}{n+1}}.
\end{equation}
Through this rescaling the field is now a function of only three variables ($\alpha$, $n$ and $\hat{\rho}$). The definition of $\alpha$ also makes it simple to discern the degeneracies of any particular model. In other words, when solving $\hat{\phi}$ for some value of $\alpha$, that solution is also valid for all combinations of $\Lambda$, $\beta$, $\rho_{0}$ and $L$ that produce the same $\alpha$-value. A similar approach to utilising degeneracies in the chameleon model was taken in Ref.~\cite{Sabulsky_2019}.

In this dimensionless rescaling, the position of the minimum of the effective potential and the Compton wavelength of fluctuations around this minimum have the simpler forms,
\begin{equation}
    \label{eq:chameleon rescaled min}
    \hat{\phi}_{\rm min}(\hat{\rho}) = \hat{\rho}^{-\frac{1}{n+1}},
\end{equation}
and
\begin{equation}
\label{eq:chameleon rescaled wavelength}
    \hat{\lambda}^2(\hat{\rho}) = \frac{\alpha }{(n+1)}\hat{\rho}^{-\frac{n+2}{n+1}}, 
\end{equation}
respectively, where $\hat{\lambda} = \lambda/L$ is the rescaled Compton wavelength.

\section{Finite element method}
\label{sec:Finite Element Method}
As discussed in the Introduction we are interested in solving for the chameleon field profile around matter distributions with complicated shapes. To do this we use the FEM to solve the chameleon equation of motion shown in equation (\ref{eq:chameleon_Norm}). Because of the ease in which this method can be adapted to arbitrary meshes, this allows us to adjust the mesh to any arbitrary source shape, and to add additional refinement where necessary. For example, for sources where the chameleon field profile exhibits the thin-shell effect, much of the variation in the field occurs at the boundaries of the dense regions. We can add additional refinement to the mesh in these regions, and make the mesh in other regions coarser to reduce computation cost whilst not sacrificing the accuracy of the solution. To perform the FEM calculations we use the FEniCS software package \cite{AlnaesBlechta2015a, LoggMardalEtAl2012a, LoggWells2010a, LoggWellsEtAl2012a}. In this section we introduce the FEM, and its application to solving the chameleon equation of motion. For a more detailed introduction to the FEM we refer the reader to Ref.~\cite{langtangen2019introduction}.

In the FEM the domain of the problem, $\Omega$, is segmented into cells, whose boundaries are defined by their vertices, $P_i$. The value of the field inside each cell is approximated by a piecewise polynomial function that matches the field values at each of the cell's vertices \cite{LangtangenLogg2017}. Extending this to the whole domain, the field, $u(\underline{x})$, can be defined using the basis functions $ e_i(\underline{x})$ as
\begin{equation}
\label{eq:phi_lin}
    u(\underline{x}) = \sum_i U_i e_i(\underline{x}),
\end{equation}
where $U_i = U(P_i)$. In this setup, $e_i$ is defined such that $e_i(P_j)=\delta_{ij}$ and between vertices $e_i$ is only nonzero in cells containing the corresponding vertex $P_i$ \cite{strang1973}. 

The FEM is designed to solve second order PDEs of the form $\nabla^2 u\left(\underline{x}\right) = -f\left(\underline{x}, u(\underline{x})\right)$, with boundary conditions $u(\underline{x}) = u_0(\underline{x})$ applied to the edge of domain, $\partial\Omega$ \cite{LangtangenLogg2017}. This is done by first transforming the second order equation into the integral of a first order equation using Green's theorem,
\begin{equation}
\label{eq:Greens_theorm}
    \int_{\Omega} \left(\nabla^2 u \right) v_j dx +  \int_{\Omega} \nabla u \cdot \nabla v_j dx = \int_{\partial\Omega} \left(\partial_n u \right) v_j dx,
\end{equation}
where $\partial_n u$ is the gradient of the field perpendicular to $\partial\Omega$. The function $v_j$ is an arbitrary test function defined to vanish on $\partial\Omega$, for all $j$ \cite{LangtangenLogg2017, langtangen2019introduction}. As a result of this choice the boundary term in equation (\ref{eq:Greens_theorm}) vanishes. Applying this boundary condition and substituting in the form of the second order PDE gives the relation
\begin{equation}
    \label{eq:FEM}
    \int_{\Omega} \nabla u \cdot \nabla v_j dx = \int_{\Omega} f(\underline{x}) v_j dx.
\end{equation}
Decomposing the field $u$ as in equation (\ref{eq:phi_lin}) we find
\begin{equation}
    \label{eq:FEM as sum}
    \sum_i \left(\int_{\Omega} \nabla e_i \cdot \nabla v_j dx \right) U_i  = \int_{\Omega} f(\underline{x}) v_j dx,
\end{equation}
which can be rewritten explicitly as a linear matrix relation,
\begin{equation}
    \label{eq:FEM matrix form}
    \mathbf{MU} = \mathbf{b},
\end{equation}
where $\mathbf{U}$ is a vector with elements $U_i$ and the matrix $\mathbf{M}$ and vector $\mathbf{b}$ are defined to be
\begin{equation}
    \label{eq:FEM in matrix form}
    \mathbf{M}_{ij} = \int_{\Omega} \nabla e_i \cdot \nabla v_j dx,
\end{equation}

\begin{equation}
    \label{eq:FEM in vector form}
    \mathbf{b}_j = \int_{\Omega} f(\underline{x}) v_j dx.
\end{equation}
The vector $\mathbf{U}$ can therefore be determined by inverting $\mathbf{M}$. The calculation of the inverse is made easier by the fact that the $e_i$ are only nonzero for cells that contain the vertex $P_i$, and so $\mathbf{M}$ is a sparse matrix \cite{strang1973}. 

\section{Nonlinear solvers}
\label{sec:Nonlinear Solvers}
Integrating the chameleon equation of motion, equation (\ref{eq:chameleon_Norm}), as described in the previous section, we find the integral form of the equation of motion:
\begin{equation}
    \label{eq:cham_norm + FEM}
    \alpha \int_{\Omega} \hat{\nabla} \hat{\phi} \cdot \hat{\nabla} v_j dx = \int_{\Omega} \hat{\phi}^{-(n+1)} v_j dx - \int_{\Omega} \hat{\rho} v_j dx.
\end{equation}
This equation is nonlinear in $\hat{\phi}$, and so a nonlinear solving method is required to compute the solution. SELCIE has two inbuilt solvers of this form, the Picard and Newton iterative solving methods. In the following subsections we outline how these solvers work and their performance. For a full discussion of these methods and proof of their validity we refer the reader to Ref.~\cite{kelley1995iterative}. 

\subsection{Picard method}
\label{subsec:Picard}
In the Picard method we take the Taylor series expansion of the potential term around some field $\hat{\phi}_k$ which is the $k^{\rm th}$ estimate of the field,
\begin{equation}
    \label{eq:potencial taylor}
    \begin{split}
    \hat{\phi}^{-(n+1)} \approx \hat{\phi}_k^{-(n+1)} - (n+1)\hat{\phi}_k^{-(n+2)}(\hat{\phi} - \hat{\phi}_k) + \mathcal{O}(\hat{\phi} - \hat{\phi}_k)^2 \\
    \approx (n+2) \hat{\phi}_k^{-(n+1)} - (n+1) \hat{\phi}_k^{-(n+2)} \hat{\phi} + \mathcal{O}(\hat{\phi} - \hat{\phi}_k)^2.
    \end{split}
\end{equation}
Neglecting higher order terms and substituting the expansion in equation (\ref{eq:potencial taylor}) into equation (\ref{eq:cham_norm + FEM}) gives the following equation for $\hat{\phi}$:
\begin{equation}
    \label{eq:picard}
    \alpha \int_{\Omega} \hat{\nabla} \hat{\phi} \cdot \hat{\nabla} v_j dx + (n+1)\int_{\Omega} \hat{\phi}_k^{-(n+2)} \hat{\phi} v_j dx = (n+2)\int_{\Omega} \hat{\phi}_k^{-(n+1)} v_j dx - \int_{\Omega} \hat{\rho} v_j dx,
\end{equation}
where we have rearranged terms so that the left-hand side of Equation (\ref{eq:picard}) is bi-linear in $\hat{\phi}$ and $v_j$, and the right-hand side is linear in $v_j$ \cite{LangtangenLogg2017}. With the equation in this form, FEniCS can then be used to solve for the field $\hat{\phi}$. We iterate this procedure by setting $\hat{\phi}_{k+1} =\hat{\phi}$ and then re-solving equation (\ref{eq:picard}) to find a new $\hat{\phi}$. We then repeat this process iteratively updating the value $\hat{\phi}_k$ each time until the condition $|\hat{\phi}_{k+1} - \hat{\phi}_k| < \delta$, where $\delta$ is the tolerance, is satisfied at all vertices. 

A downside of the Picard method outlined above is that the process gives no control over the rate of change of $\hat{\phi}$ between iterations and by extension the stability of the convergence. To address this we introduce a relaxation parameter, $\omega$, and a new update procedure so that
\begin{equation}
    \label{eq:picard update}
    \hat{\phi}_{k+1} = \omega \hat{\phi} + (1-\omega) \hat{\phi}_k.
\end{equation}
By decreasing the parameter $\omega$, the solver takes smaller step sizes between values and as such is less likely to overshoot the true solution and diverge. However, smaller step sizes also mean that the solver will take longer to converge. Therefore, some care is needed in the choice of $\omega$.

As discussed in Section \ref{sec:Finite Element Method}, the FEM calculation can be expressed as a linear equation for a vector whose elements are the values of the field at each vertex. Writing equation (\ref{eq:picard}) in this form gives \begin{equation}
    \label{eq:picard_matrix}
    \left[\alpha \mathbf{M} + (n+1)\mathbf{B}_{k}\right]\mathbf{\hat{\Phi}} = (n+2)\mathbf{C}_{k} - \mathbf{\hat{P}},
\end{equation}
where $\mathbf{\hat{\Phi}}$ is the vector whose elements are the values of the field $\hat{\phi}$ at each of the mesh vertices. Here $\mathbf{M}$ is defined as in equation (\ref{eq:FEM in matrix form}), while the matrices $\mathbf{B}_{k}$ and $\mathbf{C}_{k}$ are defined as
\begin{equation}
    \label{eq: Matrix B}
    [\mathbf{B}_k]_{ij} = \int_{\Omega} \hat{\phi}_k^{-(n+2)} e_i v_j dx,
\end{equation}
\begin{equation}
    \label{eq: Matrix C}
    [\mathbf{C}_k]_{j} = \int_{\Omega} \hat{\phi}_k^{-(n+1)} v_j dx,
\end{equation}
and $e_i$ are the basis functions of the field. The density vector $\mathbf{\hat{P}}$ is defined as
\begin{equation}
    \label{eq:FEM P def}
    \mathbf{\hat{P}}_i = \int_{\Omega} \hat{\rho}(x) v_i dx.
\end{equation}
The advantage of solving the problem in the form of equation (\ref{eq:picard_matrix}) is that since $\mathbf{M}$ and $\mathbf{\hat{P}}$ do not depend on $\hat{\phi}_k$ these can be computed once, before the iterative solver, thereby reducing the total amount of computing required.

\subsection{Newton method}
\label{subsec:Newton}
The FEM Newton iterative method solves equations of the form $F(\hat{\phi};v_j) = 0$, by evaluating the linear problem 
\begin{equation}
    \label{eq:J=-F}
    \sum_{j=1}^N \frac{\partial F}{\partial \hat{\phi}_i}\delta\hat{\phi}_i = -F(\hat{\phi}_k;v_j),
\end{equation}
where $\hat{\phi}_k$ is the $k^{\rm th}$ estimate of the field, as was the case in the Picard method, and $\delta\hat{\phi}_i$ are the elements of a correction field $\delta\hat{\phi}$ which has the same basis functions as $\hat{\phi}$ \cite{LangtangenLogg2017}. The components of the field $\hat{\phi}_k$ are then updated as
\begin{equation}
    \label{eq:Newton update}
    \hat{\phi}_{k+1} = \hat{\phi}_k + \omega \delta \hat{\phi},
\end{equation}
where again $\omega$ is a relaxation parameter. The code terminates when $|\delta \hat{\phi}| < \delta$ at all vertices, where again $\delta$ is the tolerance. For the chameleon field, we want the field $\hat{\phi}$ to satisfy
\begin{equation}
    \label{eq:F}
    F(\hat{\phi};v_j) = \alpha\int_{\Omega} \hat{\nabla} \hat{\phi} \cdot \hat{\nabla} v_j dx + \int_{\Omega}\hat{\rho} v_j dx - \int_{\Omega} \hat{\phi}^{-(n+1)} v_j dx = 0.
\end{equation}
 Therefore, equation (\ref{eq:J=-F}) becomes
\begin{equation}
    \label{eq:JF_chameleon}
    \begin{split}
       \alpha\int_{\Omega} \hat{\nabla} \delta\hat{\phi} \cdot \hat{\nabla} v_j dx + (n+1) \int_{\Omega} \hat{\phi}^{-(n+2)}_k \delta\hat{\phi} v_j dx \\ = - \alpha \int_{\Omega} \hat{\nabla} \hat{\phi_k} \cdot \hat{\nabla} v_j dx - \int_{\Omega} \hat{\rho} v_j dx + \int_{\Omega} \hat{\phi}^{-(n+1)}_k v_j dx.
    \end{split}
\end{equation}
In matrix form the equation to be solved at each iteration is
\begin{equation}
    \label{eq:JF_chameleon Matrix}
    [\alpha \mathbf{M} + (n+1) \mathbf{B}_k] \mathbf{\delta \hat{\Phi}} = -\alpha \mathbf{M} \mathbf{\hat{\Phi}}_k - \mathbf{\hat{P}} + \mathbf{C}_k,
\end{equation}
where $\mathbf{\delta \hat{\Phi}}$ and $\mathbf{\hat{\Phi}}_k$ are the vector forms of $\delta \hat{\phi}$ and $\hat{\phi}_k$ respectively. The matrices $\mathbf{M}$, $\mathbf{B}$, $\mathbf{C}$ and $\mathbf{\hat{P}}$ are defined in equations (\ref{eq:FEM in matrix form}) and (\ref{eq: Matrix B})-(\ref{eq:FEM P def}) respectively. As was the case with the Picard method, we can compute $\mathbf{M}$ and $\mathbf{\hat{P}}$ prior to the iterative solver to reduce computation time.

\subsection{Optimising solvers}
\label{subsec:Optimising Solvers}
Regardless of whether the Picard or Newton solver is chosen, when solving for $\hat{\phi}_k$ or $\delta \hat{\phi}$ in equations (\ref{eq:picard_matrix}) and (\ref{eq:JF_chameleon Matrix}) respectively, at each step it is necessary to solve a {\em linear} system of the form $\mathbf{A} \mathbf{x} = \mathbf{b} $. For large matrices $\mathbf{A}$, direct substitution is impractical, so iterative methods are required, which can be further classified into stationary and Krylov subspace methods. Stationary methods apply an operator to the residual error from some initial estimate of $\mathbf{x}$ through, for example, splitting of the matrix $\mathbf{A}$. Krylov methods work by forming a set of basis functions with successive powers of $\mathbf{A}$ applied to the residual, and are guaranteed to converge (although this may be slow for large systems). The archetypal example of a Krylov solver is the conjugate gradient (CG) method, which is suitable for symmetric positive-definitive $\mathbf{A}$. Due to the construction of the test and trial functions, the matrices in equations (\ref{eq:picard_matrix}) and (\ref{eq:JF_chameleon Matrix}) always satisfy this property \cite{LangtangenLogg2017}. The convergence of both types of iterative methods can be improved by preconditioning with a matrix $\mathbf{K}$. This involves solving the system $\mathbf{K}^{-1} \mathbf{A} \mathbf{x} = \mathbf{K}^{-1} \mathbf{b}$, where $\mathbf{K}$ is chosen such that the spectrum of eigenvalues of $\mathbf{K}^{-1} \mathbf{A}$ is close to 1, and $\mathbf{K}^{-1} \mathbf{b}$ is inexpensive to evaluate. The simplest type of preconditioner is the Jacobi (or diagonal) preconditioner, where $\mathbf{K} = \mathrm{diag} \left( \mathbf{A} \right)$.

\begin{figure}[tbp]
    \centering
    \includegraphics[width=\textwidth]{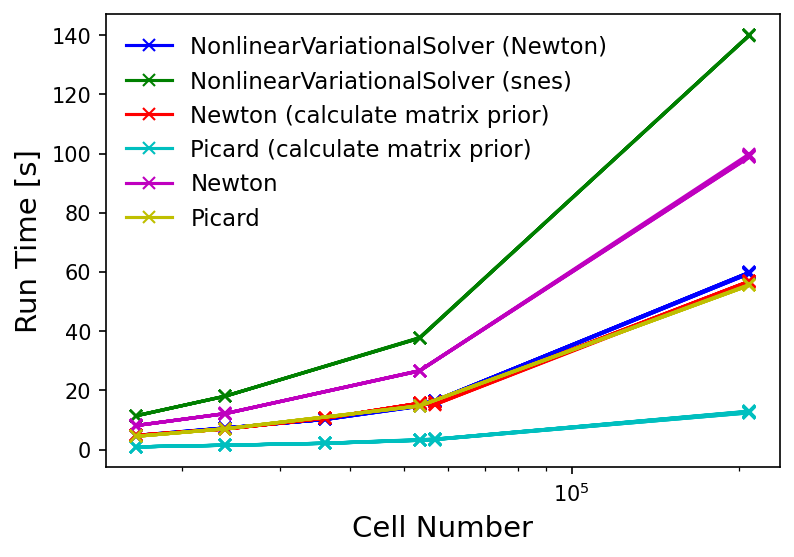}
    \caption{Plot showing how the optimum linear solver and preconditioner combinations for various nonlinear solving methods scale with mesh cell number for a 2D spherical source inside a vacuum chamber.}
    \label{fig:Optimising 2D Solver}
\end{figure}

To determine the optimal choice of nonlinear method, linear solver and preconditioner, we tested various combinations against meshes of varying cell number for a 2D spherical source inside a vacuum chamber. In total, we investigated 6 nonlinear methods: (1) Newton; (2) Picard; (3) Newton with pre-calculated system matrices; (4) Picard with pre-calculated system matrices; (5) the inbuilt FEniCS Newton solver; (6) the inbuilt FEniCS SNES solver. For each of these we tested 10 linear solvers and 11 preconditioners included in FEniCS, giving a total of 660 solver combinations. Some of these either did not work together or converge, so were excluded from the analysis. The results are summarised in Figure~\ref{fig:Optimising 2D Solver}, which shows the total run-time against cell number for each nonlinear method. In each case, we show the best combination of linear solver and preconditioner. From this, we see that the matrix form of the Picard method is both the fastest and also scales better with mesh size. In this case, the optimal combination was found to be the CG solver with Jacobi preconditioner. To check this generalised to other systems, we evaluated the solution time of the linear system for a variety of source shapes, and found only small differences. The combination of matrix Picard and CG solver with Jacobi preconditioner is therefore the default choice in SELCIE.

\section{Using SELCIE}
\label{sec:Using SELCIE}
In this section we will illustrate the typical work flow of a SELCIE user.

\subsection{Mesh generating tools}
\label{subsec:Mesh Generating Tools}
\begin{figure}[tbp]
    \centering
    \begin{subfigure}[b]{0.49\textwidth}
        \centering
         \includegraphics[width=\textwidth]{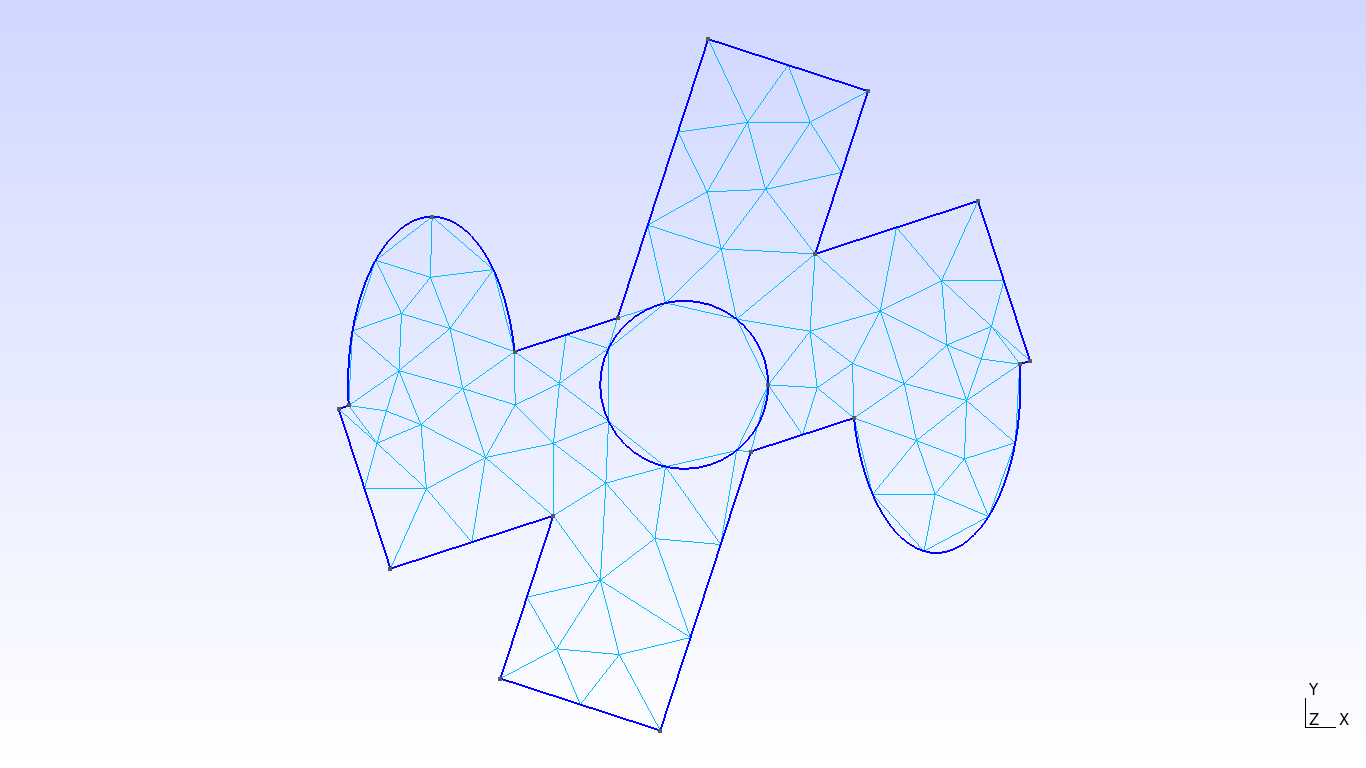}
         \caption{}
         \label{fig:Arbitary_Mesh_Example}
    \end{subfigure}
    \hfill
    \begin{subfigure}[b]{0.49\textwidth}
        \centering
         \includegraphics[width=\textwidth]{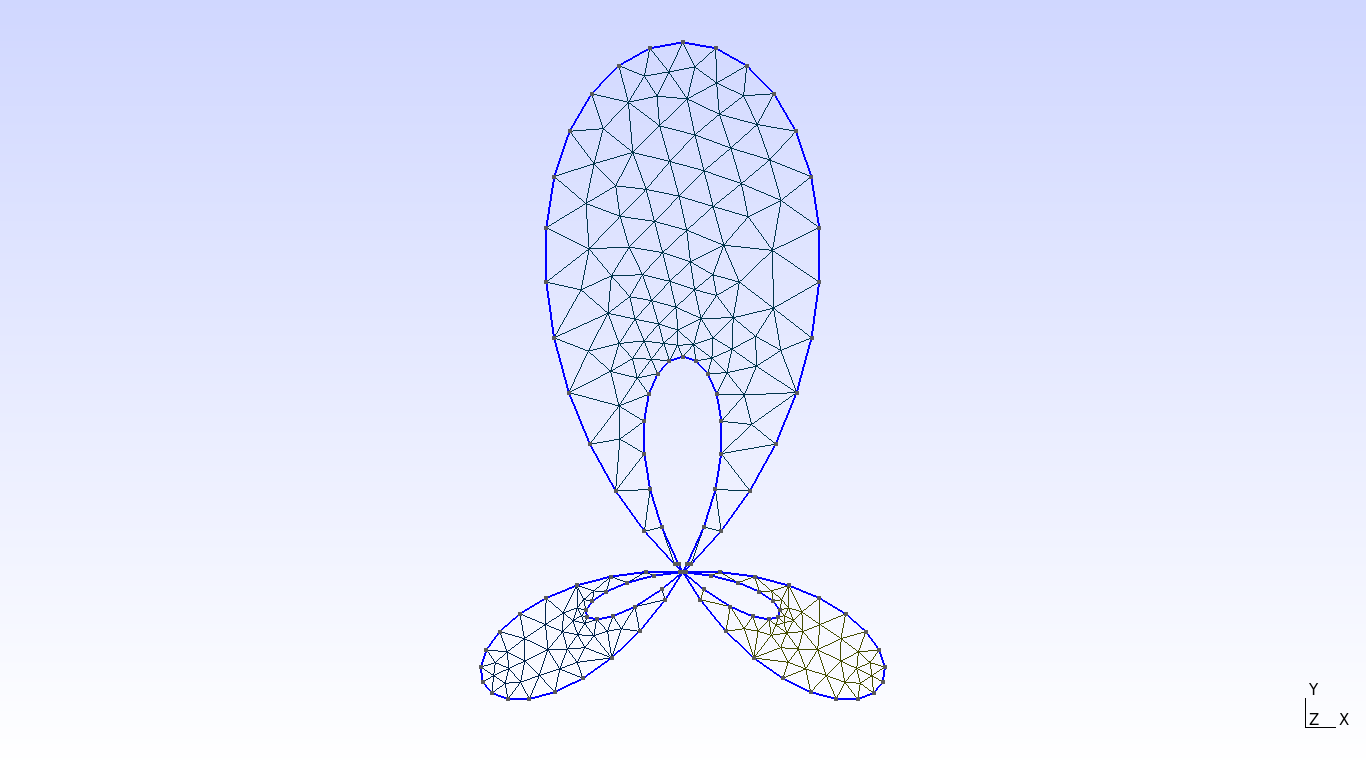}
         \caption{}
         \label{fig:Legendre_Mesh_Example}
    \end{subfigure}
    \caption{GMSH meshes constructed using: $\textbf{a)}$ a combination of functions manipulating rectangles and ellipses into a complex shape. $\textbf{b)}$ A mesh of a shape obtained from a series of Legendre polynomials with coefficients [0.0547, 0.0013, 0.0567, 0.2627], constructed using the $points\_to\_surface()$ function.}
    \label{figs:Mesh_Examples}
\end{figure}

We begin by determining the spatial regions within which we wish to solve the chameleon equations of motion, and covering them with an array of points defining a mesh. SELCIE is equipped with tools to help the user construct meshes using the GMSH software \cite{gmsh, Geuzaine2009Gmsh}. One possibility is to use the functions built into SELCIE to generate basis shapes such as ellipses and rectangles. These shapes can then be rotated, translated, combined or subtracted from one another to produce new, more complex shapes. An example of a complex shape constructed in this manner is shown in Figure \ref{fig:Arbitary_Mesh_Example}. Alternatively, it may be that the desired mesh shape is defined by a known function. In that case, SELCIE can construct the mesh directly from a list of points obtained from the function that defines the closed surface of the shape. Figure \ref{fig:Legendre_Mesh_Example} is an example of a mesh constructed using this method. These tools, either used separately or in combination, allow the user to construct a vast range of shapes without prerequisite knowledge of the GMSH interface. These shapes can also be made into subdomains of a larger mesh which can be used to evaluate the field. 

A major benefit to the finite element method is the ability to use non-uniform meshes. This can be especially useful with screened scalar field models where the field may vary more significantly in some regions of space than others. As such, SELCIE gives the user control over the size of individual cells as a function of distance from the boundaries. This can be done by defining some range for which the cell size will vary linearly between some given minimum and maximum. Once generated the mesh is saved as a $.msh$ file.

\subsection{Dimensional reduction through applying symmetry}
\label{subsec:Working in 2D}

Solving differential equations in three spatial dimensions can be computationally expensive. To reduce the number of degrees of freedom in a given problem a symmetry can be imposed, reducing the effective dimension of the problem. When applicable, this can be done in SELCIE through the introduction of a symmetry factor, $\sigma$ into the integrations,
\begin{equation}
    \label{eq:symmetry factor example}
    \int_{\Omega} f\left(\underline{x}, u(\underline{x})\right) dx \rightarrow \int_{\Omega/S} f\left(\underline{x}, u(\underline{x})\right) \sigma dx.
\end{equation}
where $S$ is the symmetry group of the applied symmetry. Systems with axis symmetry around the $x$ or $y$ axes can be simplified to 2-dimensions using the symmetry factors $|y|$ and $|x|$ respectively. The final option built into SELCIE is a translational symmetry perpendicular to the plane of the 2D mesh which has a symmetry factor equal to $1$.

\subsection{Solving for the field}
\label{subsec:Solving for the Field}

Once the mesh has been created and saved, the user can then define the matter distribution which sources the chameleon field profile. The user can define the density profile in terms of a set of functions, each defining the density on a different subdomain of the mesh. In this way, complex density profiles can be easily constructed.

After applying the appropriate symmetry factor, the user is now free to use either the $picard()$ or $newton()$ functions to solve for the field given the density profile and a choice of the chameleon parameters $n$ and $\alpha$. It is then possible to compute the field gradient (vector or scalar magnitude). 

\begin{figure}[tbp]
    \centering
    \begin{subfigure}[b]{0.49\textwidth}
        \centering
         \includegraphics[width=\textwidth]{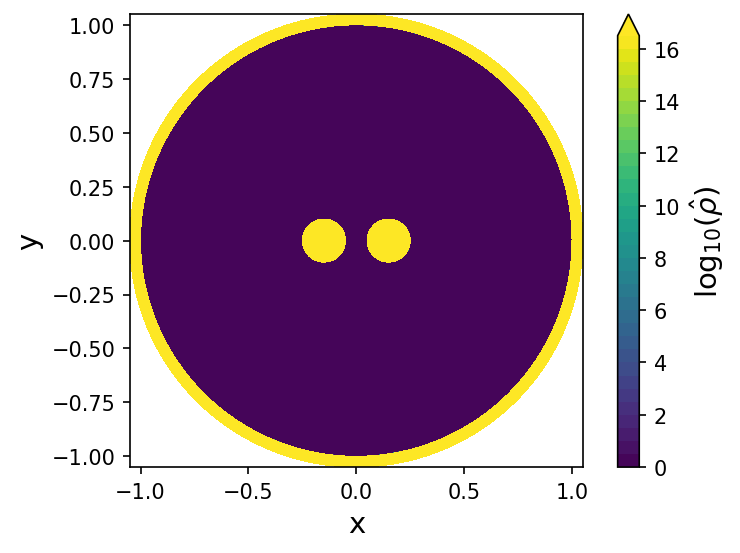}
         \caption{}
         \label{fig:Example_Subdomains}
    \end{subfigure}
    \hfill
    \begin{subfigure}[b]{0.49\textwidth}
        \centering
         \includegraphics[width=\textwidth]{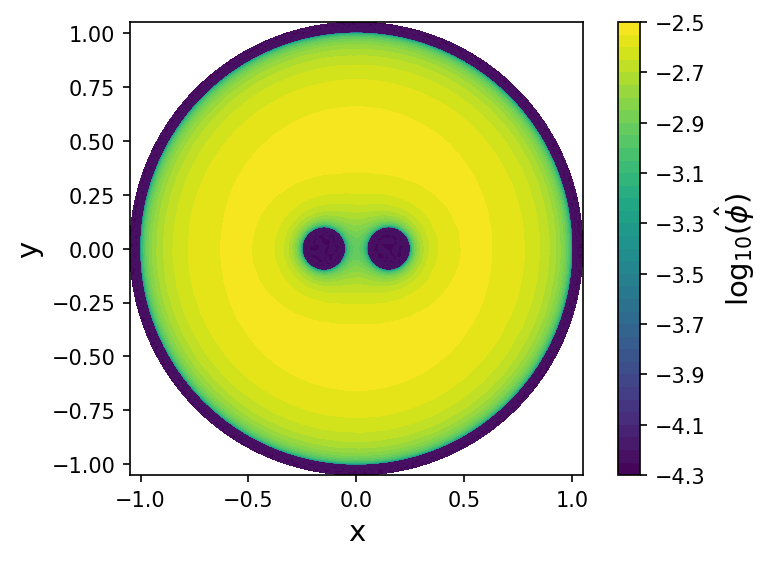}
         \caption{}
         \label{fig:Example_Field}
    \end{subfigure}
    \hfill
    \begin{subfigure}[b]{0.49\textwidth}
        \centering
         \includegraphics[width=\textwidth]{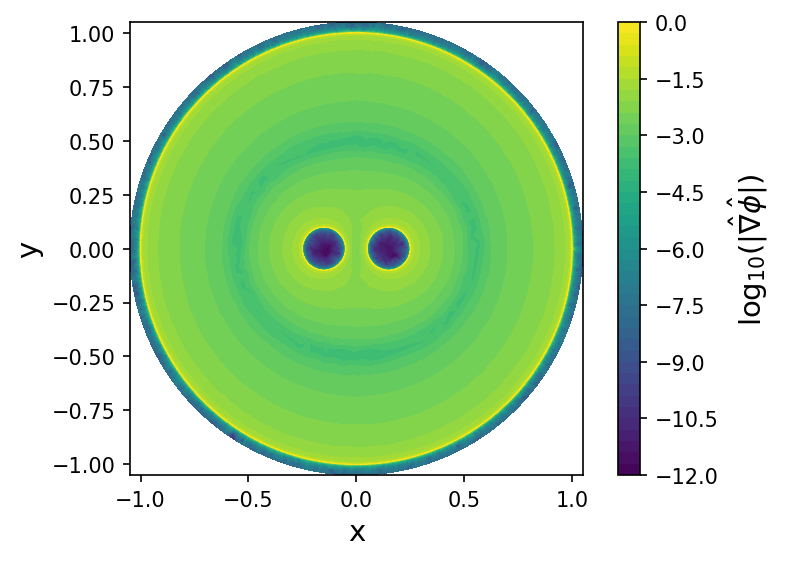}
         \caption{}
         \label{fig:Example_Grad}
    \end{subfigure}
    \hfill
        \begin{subfigure}[b]{0.49\textwidth}
        \centering
         \includegraphics[width=\textwidth]{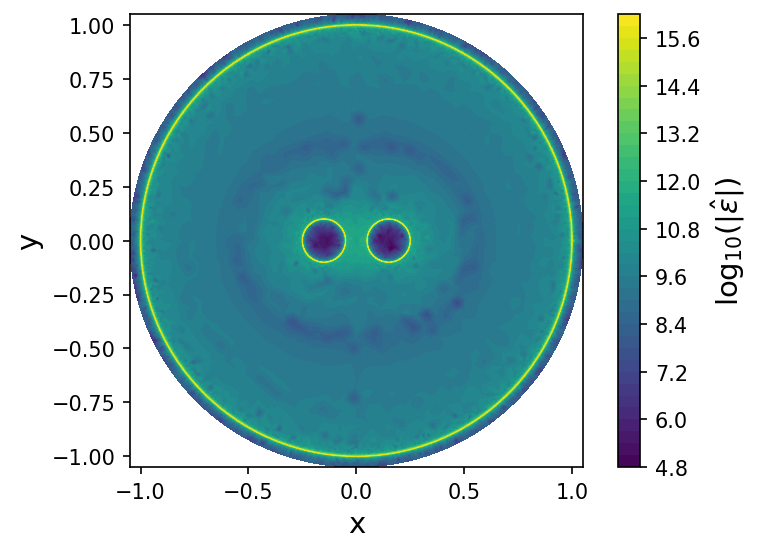}
         \caption{}
         \label{fig:Example_Res}
    \end{subfigure}
    \caption{Calculated field results from SELCIE for an axis-symmetric system in the y-axis consisting of a torus with a hole radius of $0.05$ and a tube radius of $0.1$, inside a vacuum chamber of unit radius, wall thickness $0.1$, and density equal to that of the torus. The chameleon parameters are $n=3$ and $\alpha=10^{12}$. $\textbf{a)}$ The density profile of the system. $\textbf{b)}$ The chameleon field profile. $\textbf{c)}$ The magnitude of the field gradient. $\textbf{d)}$ The strong residual of the field.}
    \label{figs:Examples}
\end{figure}

\begin{figure}
    \centering
    \begin{subfigure}[b]{0.49\textwidth}
        \centering
         \includegraphics[width=\textwidth]{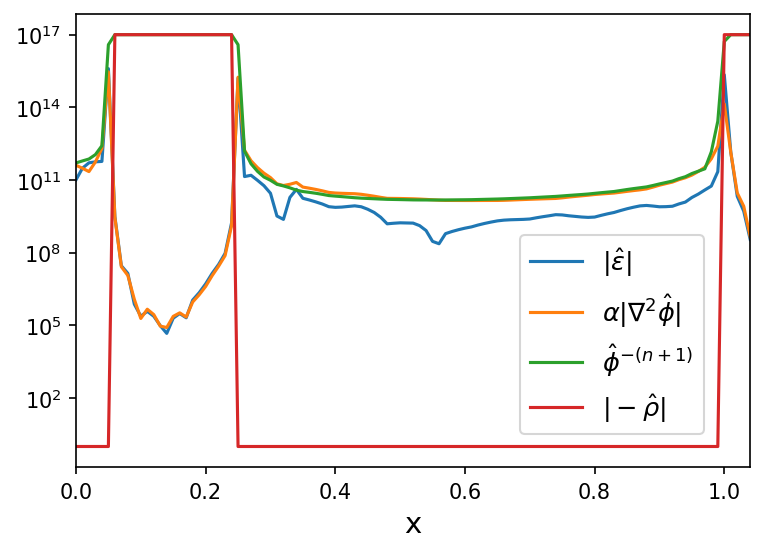}
         \caption{}
         \label{fig:Torus Residual X}
    \end{subfigure}
    \hfill
    \begin{subfigure}[b]{0.49\textwidth}
        \centering
         \includegraphics[width=\textwidth]{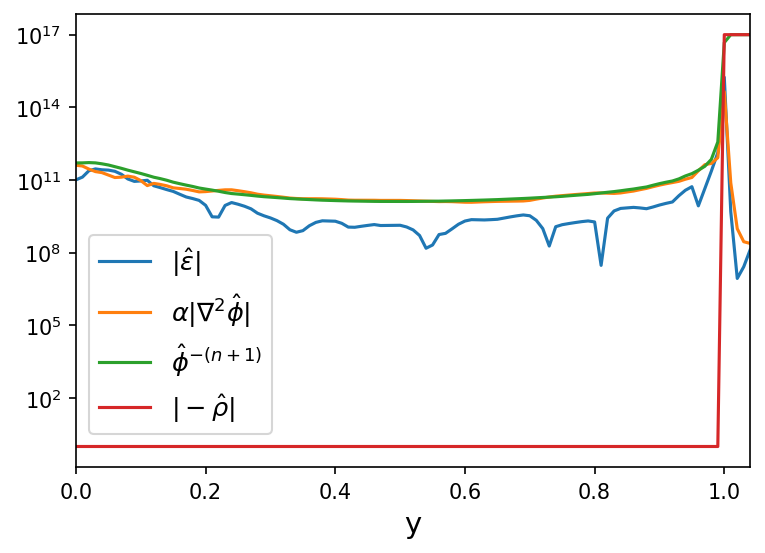}
         \caption{}
         \label{fig:Torus Residual Y}
    \end{subfigure}
    \caption{Plot of the strong residual of equation (\ref{eq:chameleon_Norm}) and the components of the equation of motion for a torus with hole radius $0.05$ and tube radius of $0.1$, inside a unit sized vacuum chamber with wall thickness $0.1$ and with axis-symmetry imposed in the y-axis. The variation of the residual and the components of the equation of motion along the lines $y=0$ and $x=0$ are shown in Figures \ref{fig:Torus Residual X} and \ref{fig:Torus Residual Y} respectively. Both the torus and the vacuum walls have a rescaled density of $10^{17}$ and the system the chameleon parameters used are $n=3$ and $\alpha=10^{12}$.}
    \label{figs:Torus Residuals}
\end{figure}

To diagnose the accuracy of the solutions to the chameleon equation of motion obtained by SELCIE, the strong residual can be evaluated. We do this by inputting the solution obtained for the scalar field into the equation of motion, equation (\ref{eq:chameleon_Norm}). The amount by which this differs from zero is the strong residual, and we say that the solutions we obtain are accurate when the strong residual is significantly smaller than the dominant term(s) in the equation of motion. As an example, Figure \ref{figs:Examples} shows the density profile of a torus inside a vacuum chamber, the associated chameleon field profile, the magnitude of the field gradient, and the strong residual. The equations have been solved assuming that the system is symmetric under rotations around the vertical $y$-axis. From Figure \ref{fig:Example_Res} it can be seen that the strong residual varies by orders of magnitude across the spatial domain. In Figure \ref{figs:Torus Residuals} we show the strong residual alongside the component terms of the equation of motion, along the $x$ and $y$-axes. In regions of high density we see that the dominant terms are $\hat{\phi}^{-(n+1)}$ and $|-\hat{\rho}|$. Meanwhile, in the vacuum regions the dominant terms are $\hat{\phi}^{-(n+1)}$ and $|\alpha \nabla^2 \hat{\phi}|$. Assuming a machine precision of $10^{-14}$ and an expected (dimensionless) field value of $\sim10^{-4}$, the expected error on the $\hat{\phi}^{-(n+1)}$ is $\sim10^{7}$. This demonstrates that the error on equation (\ref{eq:chameleon_Norm}) can be very large, even at machine precision, due to the nonlinear nature of the equation. We will consider the solutions we find to be accurate if the strong residual is at least one order of magnitude smaller than the dominant terms in the equation of motion. This can be seen to be the case in Figure \ref{figs:Torus Residuals}.

\begin{figure}[tbp]
    \centering
    \begin{subfigure}[b]{0.49\textwidth}
        \centering
         \includegraphics[width=\textwidth]{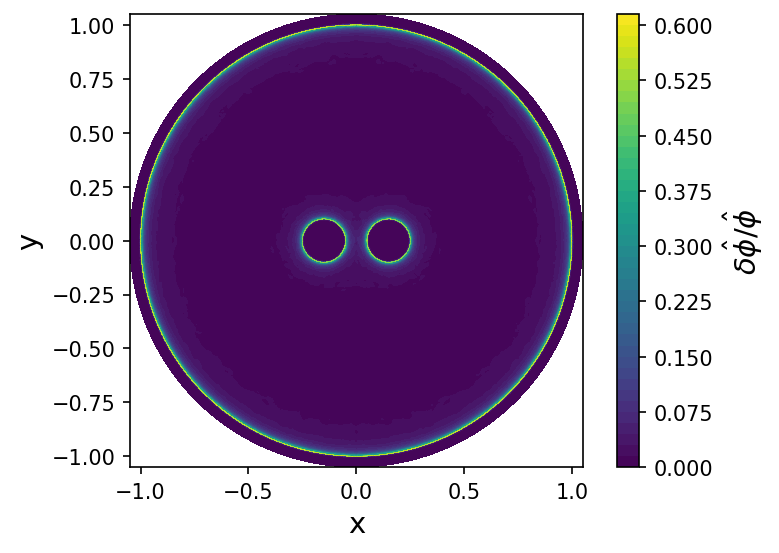}
         \caption{}
         \label{fig:Compare 2D - 3D (not zoomed)}
    \end{subfigure}
    \hfill
    \begin{subfigure}[b]{0.49\textwidth}
        \centering
         \includegraphics[width=\textwidth]{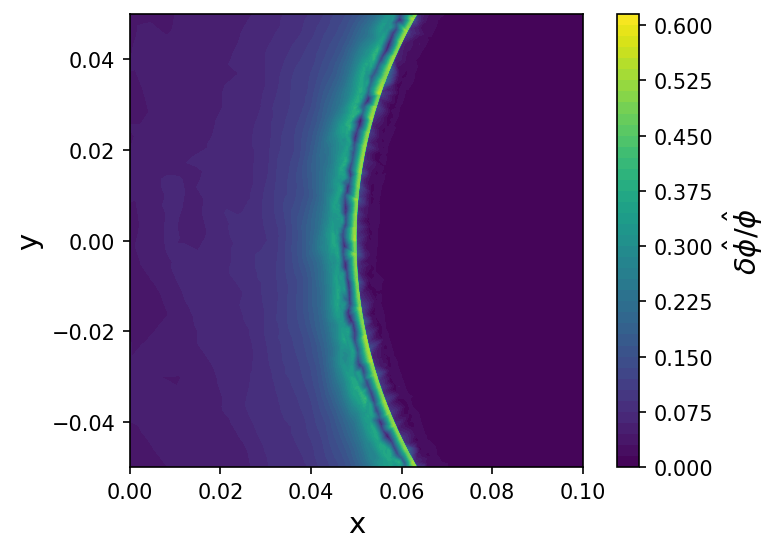}
         \caption{}
         \label{fig:Compare 2D - 3D (zoomed)}
    \end{subfigure}
    \caption{Plots of the maximum relative difference between the 2D and 3D solutions of a torus inside a vacuum chamber for a range of azimuthal angles. Figure \ref{fig:Compare 2D - 3D (not zoomed)} shows the whole domain while Figure \ref{fig:Compare 2D - 3D (zoomed)} shows a small region at the boundary of the source. The torus has a hole of radius 0.05, tube radius of 0.1 and rescaled density of $10^{17}$. The chamber wall has the same density as the torus and a thickness of $0.1$. Both measurements were measured using the chameleon parameters $n=3$ and $\alpha=10^{12}$.}
    \label{figs:Compare 2D - 3D}
\end{figure}

To check the accuracy of the solver in 3D, we can also solve for the chameleon field around the torus without imposing axis-symmetry. Figure \ref{figs:Compare 2D - 3D} shows the maximum relative difference between the 2D and 3D solutions across a range of azimuthal angles. From these plots we see a significant relative error at the discontinuous boundaries of $\sim60\%$. We found this to be a consequence of the 3D mesh not being sufficiently refined at the boundary. However, due to the scaling relation between boundary precision and cell number, it can quickly become a computationally expensive calculation to construct better refined meshes in 3D. Nevertheless, away from the boundaries the relative error decreases to percent levels and the two solutions have a strong agreement. This illustrates that even with a coarser boundary in 3D, SELCIE can still accurately determine the solution to the equation of motion.

\section{Comparison with known field profiles}
\label{sec:Comparing Code to Past Results}

In this section we will demonstrate that SELCIE can reproduce known solutions of the chameleon equation of motion. Consequently it also contains a summary of known analytic solutions and approximations to the chameleon equation of motion.



\subsection{Field maximum with no source}
\label{subsec:Field Maximum with no Source}

We will start by considering the chameleon field profile inside an empty spherical vacuum chamber. Since we define $\hat{\rho}_0 = 1$ everywhere inside the vacuum chamber, equation (\ref{eq:chameleon rescaled wavelength}) then shows that when $\alpha \gg 1$ the field's vacuum Compton wavelength is many orders larger than the size of the vacuum chamber. The field, therefore, does not have enough space to reach the value that minimises the effective potential inside the chamber and we can make the approximation $\hat{\phi}^{-(n+1)} \ll \hat{\rho}$ in the vacuum region. Applying this approximation to equation (\ref{eq:chameleon_Norm}) and defining a new field related to the original by the rescaling
\begin{equation}
    \label{eq: large alpha approx}
    \hat{\phi}(n, \hat{\rho}, \alpha) = \alpha^{-1/(n+2)} \hat{\varphi}(n, \hat{\rho}),
\end{equation}
we then find that the resulting equation for $\hat{\varphi}$,
\begin{equation}
    \label{eq:alpha independent relation}
    \hat{\nabla}^2 \hat{\varphi} = -\hat{\varphi}^{-(n+1)},
\end{equation}
is independent of $\alpha$. The equivalent relation for the $n = 1$ case was derived in Ref.~\cite{Burrage:2014oza}.
 
We see that the chameleon field profiles inside an empty vacuum chamber are all equivalent up to the rescaling in equation (\ref{eq: large alpha approx}) as long as $\alpha \gg 1$. 

It was shown in Ref.~\cite{Burrage:2014oza} that when $n=1$ the value of $\hat{\varphi}$ at the centre of an empty vacuum chamber should equal $0.69$ (to two significant figures).

\begin{figure}[tbp]
    \centering
    \includegraphics{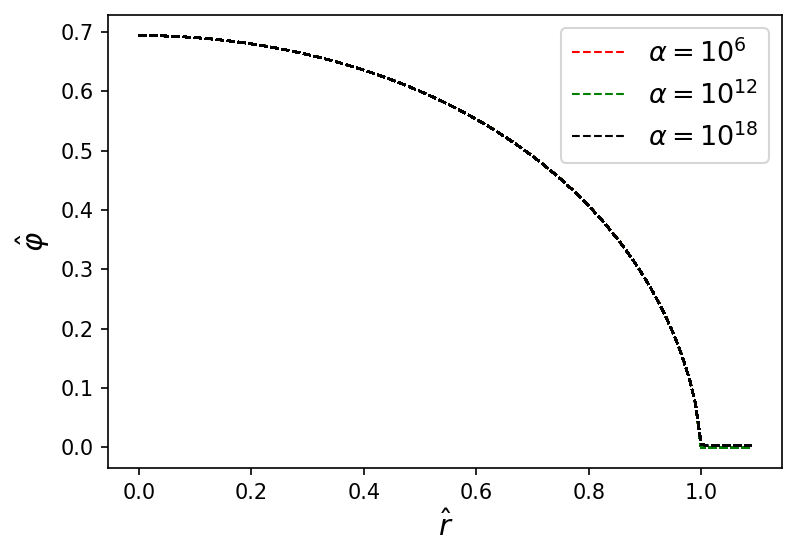}
    \caption{The rescaled chameleon field as described in equation (\ref{eq: large alpha approx}) inside an empty vacuum chamber for various values of $\alpha$. In each case $n=1$ and $\hat{\rho}_{\rm source} = 10^{17}$. Note that, except when $\hat{r}>1$, the three curves exactly overlap. }
    \label{fig:EmptyVacuum_fields}
\end{figure}

Figure \ref{fig:EmptyVacuum_fields} shows the profile of $\hat{\varphi}$ computed with SELCIE for a range of $\alpha$ values, all of which are much greater than unity. As predicted, $\hat{\varphi}$ has the same profile inside the vacuum for the whole range of $\alpha$ values tested, verifying that equation (\ref{eq:alpha independent relation}) holds. The value of $\hat{\varphi}$ at the origin was also consistent with the value found in Ref~\cite{Burrage:2014oza}.
\subsection{Solutions around circular sources}
\label{subsec:Approximate Analytical Solutions around Circular Source}

For sufficiently large and dense spherical sources (where the spherical symmetry simplifies equation ~(\ref{eq:chameleon}) to an effective 1D problem), the thin-shell solution, outlined in Section \ref{sec:The Chameleon Model}, to the chameleon equation of motion can be found by making different approximations to the chameleon equation of motion in three spatial regions. The resulting approximate solution for the exterior field profile produced by a spherical source of radius $R$ can be written as
\begin{equation}
    \label{eq:chameleon approximate solution - sphere}
    \phi(r) \approx -\left( \frac{3\beta}{4 \pi M_{\rm pl}}\right) \left( \frac{\Delta R}{R} \right) \frac{M_c e^{-m_{\rm bg} (r-R)}}{r} + \phi_{\rm bg},
\end{equation}
where $\phi_{\rm bg}$ is the value of the field that minimises the effective potential in vacuum, $m_{\rm bg}$ is the mass of the field in vacuum, $M_c$ is the mass of the source and $\Delta R$ is the width of the thin-shell \cite{Khoury:2003rn}. This expression is valid for $\Delta R/ R \ll 1$, where
\begin{equation}
    \label{eq:Del R - thin shell}
    \frac{\Delta R}{R} = \frac{|\phi_{\rm bg} - \phi_{c}|}{6 \beta M_{\rm pl} \Phi_c},
\end{equation}
$\Phi_c = M_c /8 \pi M_{\rm pl}^2 R$ is the Newtonian potential at the surface of the source and $\phi_c$ minimises the effective potential inside the source. Equation (\ref{eq:Del R - thin shell}) is the condition determining whether objects are large and dense enough (more specifically objects with a sufficiently high surface Newtonian potential) to have a thin-shell, $\Delta R/R \ll 1$, and thus be screened. The field profile around smaller sources can be found by setting $\Delta R / R = 1$ in equation~(\ref{eq:chameleon approximate solution - sphere}). 
 
Approximate analytic solutions can also be found around infinite cylindrical sources. Assuming the condition $m_{\rm bg}R \ll 1$ holds, the analytic solution to the chameleon field around a cylindrical source is
\begin{equation}
    \label{eq:chameleon approximate solution - cylinder}
    \phi(r) \approx \phi_{\rm bg} - \frac{\beta \rho_c R^2}{2 M_{\rm pl}} \left(1-\frac{S^2}{R^2}\right) \mathcal{K}_0(m_{\rm bg}r),
\end{equation}
where $\mathcal{K}_0(x)$ is a modified Bessel function \cite{Burrage:2014oza}. Here $R$ is the radius of the cylinder, $\rho_c$ is the density inside the cylinder, and $S$ is the position of the thin-shell radius given by the expression:
\begin{equation}
    \label{eq:def S - cylnder}
    \frac{4 M_{\rm pl} \phi_{\rm bg}}{\beta \rho_c R^2} = \left(1-\frac{S^2}{R^2}\right)\left(1 - 2 \gamma_E - 2 \ln\left(\frac{m_{\rm bg}R}{2}\right) \right) + \frac{2 S^2}{R^2} \ln\left(\frac{S}{R} \right),
\end{equation}
where $\gamma_E$ is the Euler-Mascheroni constant \cite{Burrage:2014oza}. We see that only sufficiently large sources have a non-trivial solution to equation (\ref{eq:def S - cylnder}) and therefore a thin-shell, as was true in the spherical case.

For ease of comparison with the results of SELCIE we state here the rescaled form of the spherical and cylindrical solutions, in the sense of Section \ref{sec:The Chameleon Model}. Equation (\ref{eq:Del R - thin shell}) describing the position of the thin-shell inside a spherical object becomes
\begin{equation}
    \label{eq:Del R - thin shell - rescaled}
    \frac{\Delta \hat{R}}{\hat{R}} = \frac{\alpha (1 - \hat{\phi}_c)}{\hat{\rho}_c \hat{R}^2},
\end{equation}
where $R$ and $\Delta R$ have been rescaled to $\hat{R}=R/L$ and $\Delta \hat{R}=\Delta R /L$, respectively. Assuming $\hat{\rho}_c \gg 1$, from equation (\ref{eq:chameleon rescaled min}) we see that $\hat{\phi}_c \ll 1$. Applying this to equation (\ref{eq:Del R - thin shell - rescaled}) and substituting it into a rescaled equation (\ref{eq:chameleon approximate solution - sphere}) gives the rescaled field profile for a spherical source:
\begin{equation}
    \label{eq:chameleon approximate solution - sphere - rescaled}
    \hat{\phi}(\hat{r}) \approx 1 - \frac{\hat{R}}{\hat{r}} e^{-(\hat{r} - \hat{R})\sqrt{\frac{(n+1)}{\alpha}}}.
\end{equation}

For the cylindrical solution, assuming $(R-S)/R \ll 1$, in equation (\ref{eq:def S - cylnder}) and substituting this into a rescaled equation (\ref{eq:chameleon approximate solution - cylinder}) gives the relation
\begin{equation}
    \label{eq:chameleon approximate solution - cylinder - rescaled}
    \hat{\phi} \approx 1 - \frac{2 \mathcal{K}_0 \left( \hat{r} \sqrt{\frac{(n+1)}{\alpha}} \right)}{\ln \left(\frac{4 \alpha}{(n+1) \hat{R}^2}\right)}.
\end{equation}
As before, this relation holds for $m_{\rm bg}R \ll 1$ which is equivalent to $(n+1)\hat{R}^2 \ll \alpha$.

\begin{figure}[tbp]
    \centering
    \includegraphics[width=\textwidth]{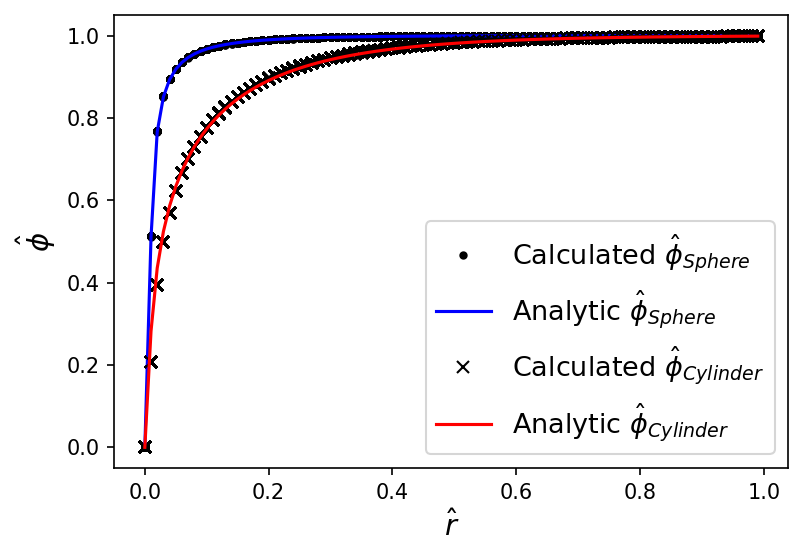}
    \caption{The calculated field profiles for the spherical and cylindrical cases against the approximate analytic solutions. For both cases, $n=1$, $\alpha=0.1$, $\hat{\rho}_{source}=10^{17}$, and the source radius was set to 0.005.}
    \label{figs:compare_sphere_cylinder}
    \label{fig:Circle_in_Chamber_compare_fields}
\end{figure}

To verify that SELCIE can reproduce these approximate analytic solutions, we constructed a 2D mesh of a circular source of radius $\hat{R} = 0.005$ inside a circular vacuum chamber of radius unity. Both spherical and cylindrical cases can be explored using the above 2D mesh by imposing axis-symmetry (along either the $x$ or $y$-axis) and translational symmetry normal to the mesh, respectively. For the values of the corresponding symmetry factors see Section \ref{subsec:Working in 2D}. In both cases we choose $n = 1$, $\alpha = 0.1$ and the source density was set to $\hat{\rho}_{c} = 10^{17}$. The analytic and numerical field profiles for both spherical and cylindrical sources are shown in Figure \ref{fig:Circle_in_Chamber_compare_fields}. From this plot we see that SELCIE is able to reproduce the analytic results to within the accuracy of the analytic approximations.

\subsection{Solutions around ellipsoidal sources}
\label{subsec:Approximate Analytical Solutions around Elliptical Source}
In Ref.~\cite{Burrage_2015} a similar approach to that outlined in Section \ref{subsec:Approximate Analytical Solutions around Circular Source} was taken to determine an approximate analytic solution for the chameleon field around an ellipsoidal source using the coordinate system:
\begin{equation}
    \begin{split}
        x &= a \sqrt{(\xi^2 - 1)(1 - \eta^2)} \cos{(\phi)}, \\
        y &= a \sqrt{(\xi^2 - 1)(1 - \eta^2)} \sin{(\phi)}, \\
        z &= a \xi \eta.
    \end{split}
    \label{eq:Ellispe coordinates}
\end{equation}
In these coordinates, $a$ is the distance from the ellipse's foci to the origin, $\eta$ is analogous to an angular coordinate ($-1 \leq \eta \leq + 1$), $\xi$ is analogous to a radial coordinate ($1 \leq \xi < \infty$), and $\phi$ is the azimuthal angle ($0 \leq \phi \leq 2 \pi$). Outside a screened ellipsoidal source the approximate analytic solution for a chameleon field is
\begin{equation}
    \phi(\xi, \eta) = \phi_{\rm bg} - \frac{a^2 \beta \rho_c}{3 M_{\rm pl}} \big[\xi_{0} (\xi_{0}^2 - 1) - \xi_{c} (\xi_{c}^2 - 1)\big] \big(Q_0(\xi) - P_2(\eta) Q_2(\xi)\big),
    \label{eq:chameleon approximate solution - ellipse}
\end{equation}
where $\xi_0$ defines the surface of the ellipse, $\phi_{\rm bg}$ is the background value of the field and $P_i$ $\&$ $Q_j$ are Legendre functions of the first and second kind respectively. The position of the interior surface of the thin-shell $\xi_c$ is defined by:
\begin{equation}
    \frac{6 M_{\rm Pl}}{a^2 \beta \rho_c} \phi_{\rm bg} + (\xi_{c}^2 - 1)\big\{1 + 2 \xi_c Q_0(\xi_c)\big\} = (\xi_{0}^2 - 1)\big\{1 + 2 \xi_0 Q_0(\xi_0)\big\}.
    \label{eq:Del xi - thin shell}
\end{equation}
Making the approximation $\delta \xi = \xi_0 - \xi_c \ll \xi_0$ and rescaling the equations as described in Section \ref{sec:The Chameleon Model}, this constraint becomes
\begin{equation}
    \delta \xi = \frac{3 \alpha \hat{\phi}_{\rm bg}}{\hat{a}^2 (3 \xi_{0}^2 - 1) Q_0(\xi_0)} \ll \xi_0,
    \label{eq:Del xi - thin shell - rescaled}
\end{equation}
where $\hat{a} = a/L$, 
and equation (\ref{eq:chameleon approximate solution - ellipse}) becomes
\begin{equation}
    \hat{\phi}(\xi, \eta) = \hat{\phi}_{\rm bg} \left(1 - \frac{Q_0(\xi) - P_2(\eta) Q_2(\xi)}{Q_0(\xi_0)} \right).
    \label{eq:chameleon approximate solution - ellipse - rescaled}
\end{equation}

\begin{figure}[tbp]
    \centering
    \begin{subfigure}[b]{0.49\textwidth}
        \centering
         \includegraphics[width=\textwidth]{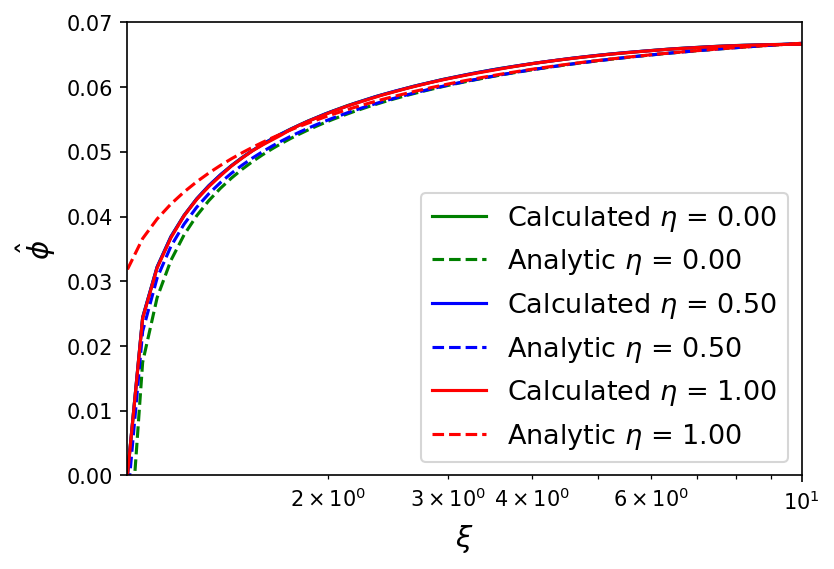}
         \caption{$\xi_0 = 1.01$}
         \label{fig:Ellipse Xi0 = 1.01}
    \end{subfigure}
    \hfill
    \begin{subfigure}[b]{0.49\textwidth}
        \centering
         \includegraphics[width=\textwidth]{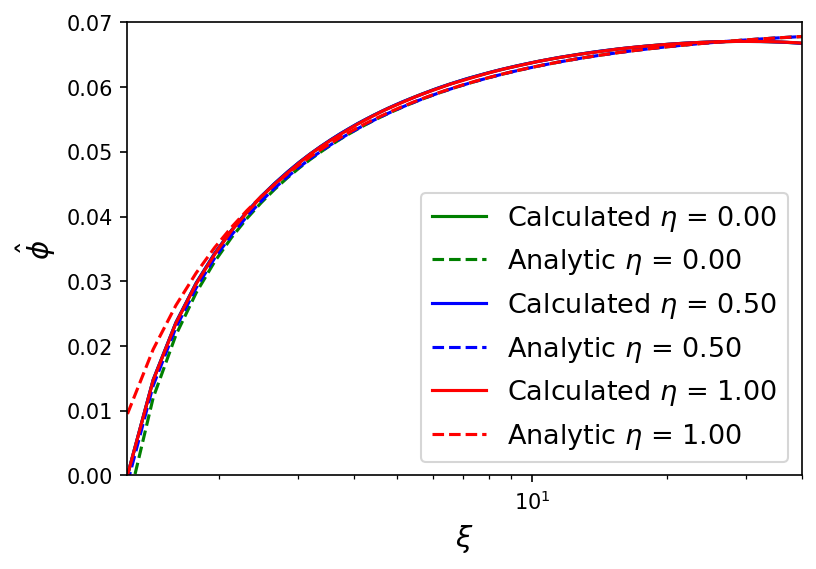}
         \caption{$\xi_0 = 1.25$}
         \label{fig:Ellipse Xi0 = 1.25}
    \end{subfigure}
    \hfill
    \begin{subfigure}[b]{0.49\textwidth}
        \centering
         \includegraphics[width=\textwidth]{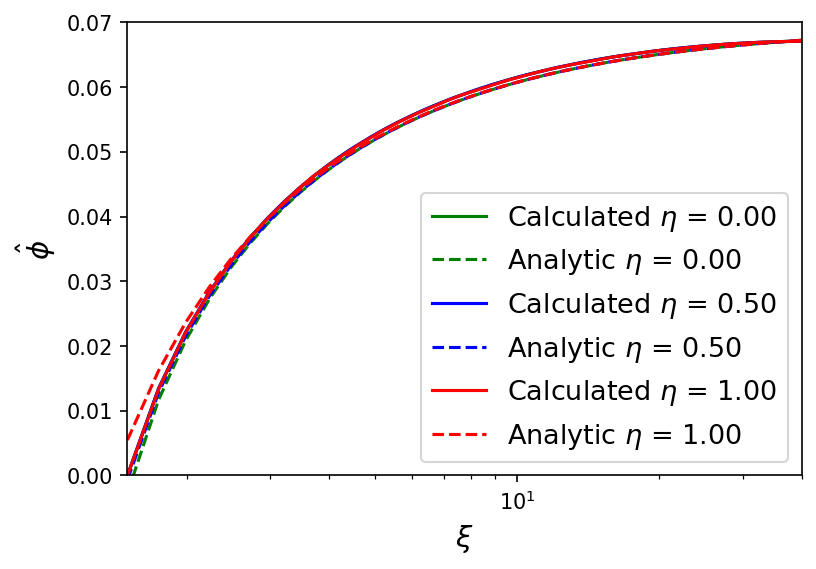}
         \caption{$\xi_0 = 1.50$}
         \label{fig:Ellipse Xi0 = 1.50}
    \end{subfigure}
    \hfill
        \begin{subfigure}[b]{0.49\textwidth}
        \centering
         \includegraphics[width=\textwidth]{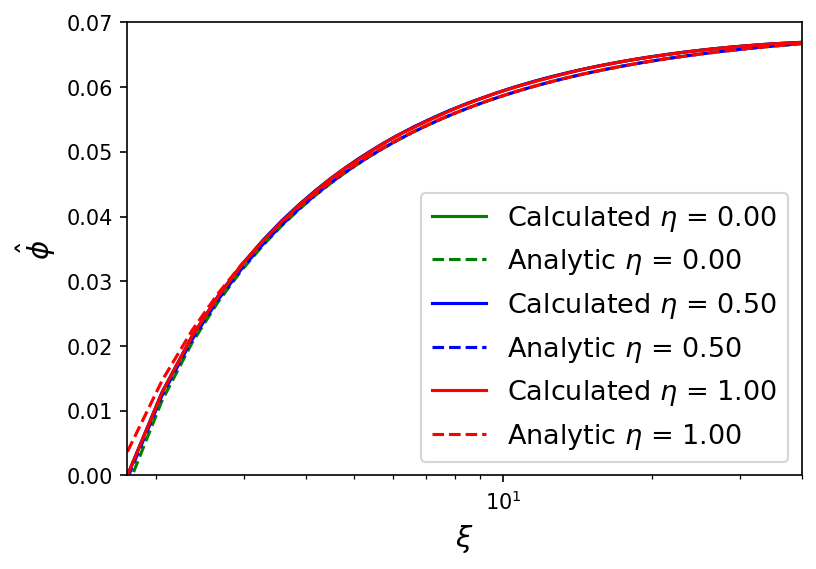}
         \caption{$\xi_0 = 1.75$}
         \label{fig:Ellipse Xi = 1.75}
    \end{subfigure}
    \caption{Comparison between the chameleon field profiles calculated by SELCIE around an elliptical source of uniform density and the approximate analytic solution shown in equation (\ref{eq:chameleon approximate solution - ellipse - rescaled}) for various values of $\xi_0$. The parameters used were $n = 1$ and $\alpha = 10^3.$}
    \label{figs:Compare Ellipses}
\end{figure}

To test this solution against the field profile found by SELCIE around an elipsoidal source, we first constructed a mesh of an ellipse inside a unit radius vacuum chamber with a wall thickness of $0.1$. The value of $\xi_0$ for this ellipse was allowed to vary while the value of $a$ was set such that for any $\xi_0$ the volume was set to that of a sphere of radius $0.005$. Taking $\rho_c = 10^{17}$, $n = 1$ and $\alpha = 10^{3}$ the chameleon field inside the chamber was calculated using SELCIE for various values of $\xi_0$. These results are compared to equation (\ref{eq:chameleon approximate solution - ellipse - rescaled}) in Figure \ref{figs:Compare Ellipses}. Close to the surface the analytic result appears to break down as the value of the field is negative. Away from the surface, however, the results from SELCIE agree with the analytic results with $\sim 1\%$ relative error. Given that this is only an approximate analytic solution, we consider our results to be in good agreement.


\subsection{Analytic solutions for NFW galaxy cluster halos}
\label{subsec:NFW}

In the previous subsections we have discussed solutions to the chameleon equation of motion around dense sources, with clearly defined surfaces, inside a vacuum chamber. In this section we will demonstrate that SELCIE can also provide accurate chameleon field profiles for continuous density distributions. An important example of such a system, of relevance to a number of observational tests, is galaxy clusters. These are the largest gravitationally bound systems in the universe. The ability to measure cluster masses in a variety of ways (X-ray surface brightness, the Sunyaev–Zeldovich effect, weak lensing and other methods \cite{Ettori_2013, Applegate_2016, Lyapin_2019}) makes clusters invaluable when studying and testing gravity on cosmological scales. In addition, galaxy clusters are known to have a complicated density distribution, which will need to be evaluated when testing for screened models such as the chameleon. These features have been employed in astrophysical fifth force searches, producing strong constraints on chameleon gravity and other models \cite{Terukina2014, Wilcox2015, Sakstein2016, Koyama2016, Burrage2018, Cataneo2018, Tamosiunas2019,tamosiunas2021chameleon}.

It has been shown that the underlying density distribution on galaxy and galaxy cluster scales is well-approximated by the Navarro–Frenk–White (NFW) profile \cite{Navarro1997}. After rescaling, this profile can be written as 

\begin{equation}
\hat{\rho}(r)=\frac{\hat{\rho}_{s}}{\hat{r}\left(1+\hat{r}\right)^{2}},
\label{NFW_profile}
\end{equation}

\noindent where the radial coordinate has been rescaled by the scale radius of the cluster $r_s$, such that $\hat{r} = r/ r_{s}$, and the density has been rescaled by the cosmological critical density $\rho_c$ such that $\hat{\rho} = \rho/ \rho_c$ and $\hat{\rho}_s = \rho_s/ \rho_c$. The NFW profile of equation (\ref{NFW_profile}) diverges as $\hat{r} \rightarrow 0$, so we also introduce a core radius, $\hat{r}_{\rm cut}$, so that for $\hat{\rho} \left(\hat{r} < \hat{r}_{\rm cut}\right) = \hat{\rho} \left(\hat{r}_{\rm cut} \right)$.

There is no exact analytical solution for a chameleon field profile within and around a spherical NFW halo. However, as in the previous subsections, an approximate analytic solution can be obtained by using a piecewise approach as shown in Ref.~\cite{Terukina2014}. Applying the rescaling outlined in Section \ref{sec:The Chameleon Model}, this solution can be written as 
\begin{equation}
    \label{eq:NFW analyitic solution - rescaled}
    \hat{\phi}(\hat{r}) = \begin{dcases}
    \hat{\phi}_s \left[\hat{r} \left(1+\hat{r}\right)^2\right]^{\frac{1}{n+1}} & \hat{r} \leq \hat{r}_c \\
    \hat{\phi}_{0} \left(1 - \frac{\hat{r}_c}{\hat{r}} \right) + \frac{\hat{\rho}_s}{\alpha} \frac{1}{\hat{r}} \ln{\left( \frac{1 + \hat{r}_c}{1 + \hat{r}} \right)} & \hat{r} \geq \hat{r}_c,
    \end{dcases}
\end{equation}
where $\hat{\phi}_s = \hat{\phi}_{\rm min}(\hat{\rho}_s)$, as defined by equation (\ref{eq:chameleon rescaled min}), $\hat{\phi}_{0}$ is the field value at spatial infinity (or at the boundary of the numerical simulation), and $\hat{r}_c$ is the transition scale. For distances less than $\hat{r}_c$ the potential and matter terms dominate over the gradient term and so the field takes the value that minimises the effective potential as given by equation (\ref{eq:chameleon rescaled min}). For scales larger than $\hat{r}_c$ the field takes values away from this minimum as the potential term becomes subdominant to the gradient and matter terms. The transition scale is $\hat{r}_c \approx (\hat{\rho}_s/\alpha \hat{\phi}_{0}) - 1$ where, in both this expression and equation (\ref{eq:NFW analyitic solution - rescaled}), it has been assumed that $\hat{\rho}_s \gg 1$ and therefore $\hat{\phi}_s \ll 1$.

\begin{figure}[tbp]
    \centering
    \begin{subfigure}[b]{0.49\textwidth}
        \centering
         \includegraphics[width=\textwidth]{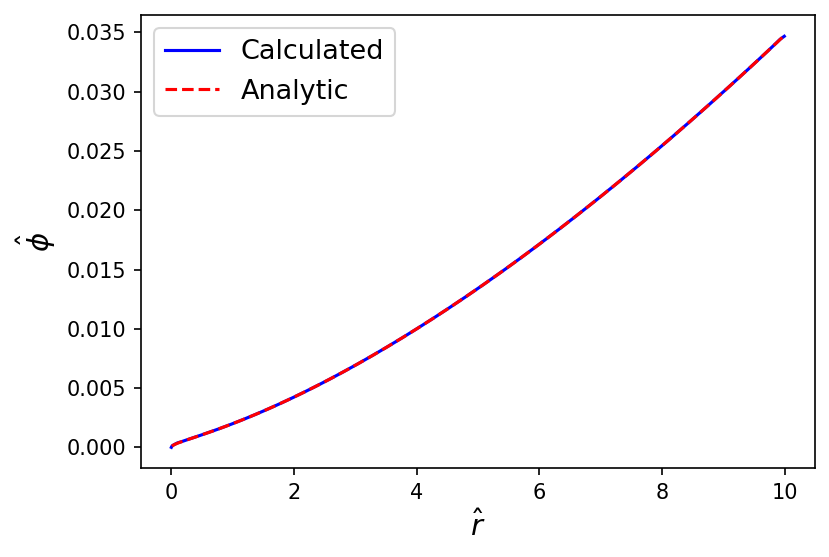}
         \caption{}
         \label{fig:NFW - small alpha}
    \end{subfigure}
    \hfill
    \begin{subfigure}[b]{0.49\textwidth}
        \centering
         \includegraphics[width=\textwidth]{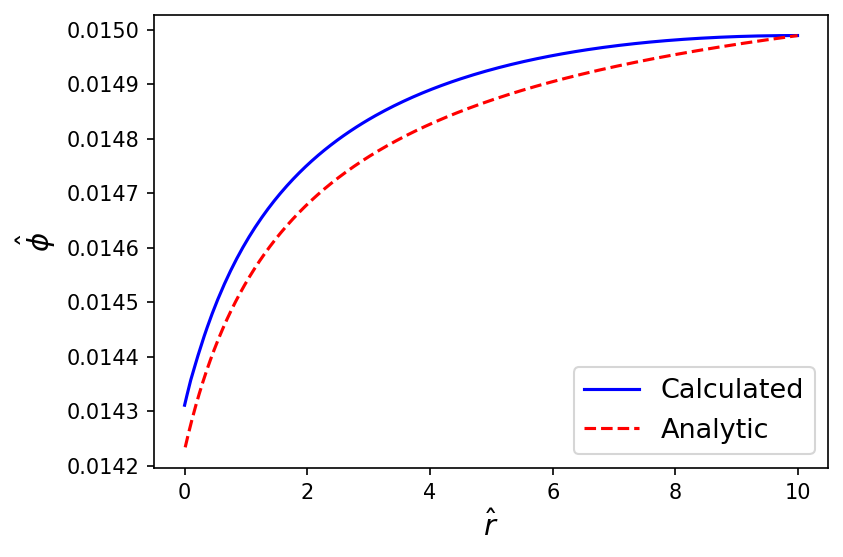}
         \caption{}
         \label{fig:NFW - large alpha}
    \end{subfigure}
    \caption{Plots comparing the field profile calculated by SELCIE with the expected analytic solution in the regime: $\textbf{a)}$ $r_c \gg 1$ $\left(\alpha = 10^{-9}\right)$, $\textbf{b)}$ $r_c \ll 1$ $\left(\alpha = 10^{9}\right)$. In both cases the rescaled critical density was set to $\rho_s = 10^{6}$.}
    \label{figs:NFW}
\end{figure}

We compared equation (\ref{eq:NFW analyitic solution - rescaled}) against solutions calculated using SELCIE for the cases when $r_c$ is much larger than the domain size and for $\hat{r}_c \ll 1$. These results are plotted in Figure \ref{figs:NFW}. In both cases the calculated and analytic results have a strong agreement, with sub-percentage relative error. However, we should mention here that due to the before mentioned cutoff that was introduced to the density profile, the assumption that the field traces the minimum of the effective potential near the origin can be broken. This can be seen in Figure \ref{fig:NFW - large alpha} where the analytic solution predicts the field would tend to zero at the origin but the calculated field tends to some larger value instead.



\subsection{Solutions around Legendre polynomial shapes}
\label{subsec:Legendre Polynomial Shapes}

As another comparison to work in the literature, we attempted to reproduce the results from Ref.~\cite{Burrage:2017shh}. This paper attempted to use a Picard nonlinear solving method with the FEM to solve the chameleon field around irregularly shaped sources. 
In this sense, SELCIE can be viewed as a continuation of this work. Ref.~\cite{Burrage:2017shh} aimed to investigate what shapes would optimise experimental searches for the chameleon field by maximising the resulting fifth force. Specifically, the shapes tested were constructed using Legendre functions, $P_i \left( \cos{\theta} \right)$, as a basis such that shapes are defined
\begin{equation}
    \label{eq:Legendre polynomial def}
    R \left(\cos{\theta}\right) = \sum_{i=0}^{3} a_i P_i \left( \cos{\theta} \right),
\end{equation}
where $a_i$ are the coefficients of the shape. The shapes were placed in a vacuum chamber of radius $15 \mbox{ cm}$ and wall thickness of $3 \mbox{ cm}$s. The fifth force was measured for a distance of $5 \mbox{ mm}$s from the surface of the source. The wall and source density were both set to $1 \mbox{ g/cm}^3$ while the vacuum density was $10^{-17} \mbox{ g/cm}^3$. The remaining parameters used were: $n = 1$, $M = M_{\rm pl}/\beta = 10^{18} \mbox{ GeV}$ and $\Lambda = 10^{-12} \mbox{ GeV}$. Using SELCIE we reconstructed this setup for each Legendre polynomial shape tested in Ref.~\cite{Burrage:2017shh}. The densities were rescaled by the vacuum value stated above and the length scale $L$ was set to the chamber radius. The value of $\alpha$ was calculated using equation (\ref{eq:alpha_def}) to be $6.1158\times 10^{18}$. Our results are presented alongside the original results from Ref.~\cite{Burrage:2017shh} in Table \ref{tab:Comparison to Legendre Shapes}.

From this table we see that our results are in disagreement with the previous results. In fact, both the individual values and the increase in the fifth force from the spherical case (bottom row) disagree. However, due to the extensive testing we have performed on SELCIE, we are confident in its results and believe there is an error in the calculation of the results reported in Ref.~\cite{Burrage:2017shh}. The new results for these shapes, found with SELCIE, could be interpreted as implying that optimising the shape may not yield as large an increase to the fifth force as previously suggested in Ref.~\cite{Burrage:2017shh}. However, since we are not confident in the previous results, we also have no reason to believe that the Legendre coefficients listed are the optimum values. It is still possible that for some choice of coefficients there is a significant increase to the fifth force compared to the spherical case.

\begin{table}[tbp]
    \centering
     \def\arraystretch{2}
    \begin{tabular}{ |c|c|c|c|c|c| } 
        \hline
        $a_0$ & $a_1$ & $a_2$ & $a_3$ & $\displaystyle{\left.\frac{\delta g}{g}\right|_{P} [10^{-11}]}$ &$\displaystyle{ \left.\frac{\delta g}{g}\right|_{S} [10^{-11}]}$ \\[1.5ex]
        \hline
        0.82 & 0.02 & 0.85 & 3.94 & 4.82 & 2.25  \\ 
        0.97 & 0.59 & 0.03 & 3.99 & 4.68 & 2.24  \\
        1.34 & 0.18 & 0.41 & 2.89 & 4.52 & 2.26  \\
        1.47 & 0.19 & 0.27 & 2.63 & 4.47 & 2.29  \\
        2.00 & 0.00 & 0.00 & 0.00 & 1.73 & 2.24  \\
        \hline
    \end{tabular}
    \caption{Table of fifth forces measured from Ref.~\cite{Burrage_2015} (labelled as $P$) compared to results produced from SELCIE (labelled as $S$). The $a_i$ values correspond to the Legendre coefficients used to define the shapes in \cite{Burrage_2015}.}
    \label{tab:Comparison to Legendre Shapes}
\end{table}




\section{Conclusion}
\label{sec:Conclusion}
In this work we have introduced SELCIE as a tool to help study the chameleon field. Currently SELCIE solves the static chameleon equation of motion in the form of equation (\ref{eq:chameleon_Norm}). Using SELCIE the chameleon field equation can be solved for highly non-symmetric systems constructed by the user. This can therefore be used to find solutions for systems which lack an analytic solution and to optimise existing experiments used to detect the chameleon field. 

We have demonstrated the reliability of SELCIE to reproduce approximate analytic results from the literature for a variety of sources and density profiles; however, its functionality is much broader than this. For example, density distributions from N-body simulations or galaxy halo profiles can be easily inputted into SELCIE to find the corresponding chameleon profile and fifth force, enabling precision tests of the chameleon model on cosmological scales.

In a separate work, Ref.~\cite{tamosiunas2021chameleon}, SELCIE has already been used to test the viability of detecting a chameleon signal through observations of galaxy cluster halos. In future work we plan to use SELCIE with a genetic algorithm to find experimental configurations which optimise the possibility of detecting a chameleon fifth force. This optimisation algorithm could be used, for example, to extend the reach of current and future experiments that search for chameleon fifth forces with atom interferometry \cite{Burrage:2014oza,Hamilton:2015zga,Elder:2016yxm,Jaffe:2016fsh,Sabulsky:2018jma}, ultra-cold neutrons \cite{Brax:2014gja,Jenke:2014yel,Lemmel:2015kwa,Li:2016tux,Sponar:2021rvk,Jenke:2020obe}, torsion balances \cite{Upadhye:2012qu,Pernot-Borras:2021edr,Zhao:2021anp}, Casimir experiments \cite{Decca:2007jq,Almasi:2015zpa,Sedmik:2018kqt,Klimchitskaya:2021lak,Sedmik:2021iaw} or opto-mechanical sensors \cite{Geraci:2010ft,Rider:2016xaq,Liu:2018jia,Qvarfort:2021zrl}.

We intend to continue to develop and extend SELCIE in the future to generalise the methodology to work with alternative forms of the chameleon model and other screened scalar fields such as the symmetron. We also plan to extend the code to allow the density profile and the field to evolve in time. Throughout the literature the chameleon field is assumed static,\footnote{Notable exceptions include Refs. \cite{Silvestri:2011ch,Sakstein:2013pda,Upadhye:2013nfa,Brax:2013uh,Hagala:2016fks,Ikeda:2021pkb}.} and so it would be of interest to study what effects a dynamic system has on the field. This could lead to strengthening constraints by re-evaluating known systems (e.g. Earth-Moon system) or by developing new experiments that utilise dynamical systems.

\acknowledgments

We would like to thank Daniela Saadeh for helpful discussions and advice. We would also like to thank the greater FEniCS community, whos work in creating FEniCS is what allowed this work to be possible. 

Clare Burrage and Andrius Tamosiunas are supported by a Research Leadership Award from The Leverhulme Trust. Chad Briddon is supported by the University of Nottingham. Adam Moss is supported by a Royal Society University Research Fellowship.

\bibliographystyle{JHEP-custom}
\bibliography{References.bib}{}

\providecommand{\href}[2]{#2}\begingroup\raggedright\begin{thebibliography}{10}

\bibitem{Aad_2012}
G.~Aad, T.~Abajyan, B.~Abbott, et~al., {\it Observation of a new particle in
  the search for the standard model higgs boson with the atlas detector at the
  lhc},  {\em Physics Letters B} {\bf 716} (Sep, 2012) 1–29.

\bibitem{Riess:1998cb}
{\bf Supernova Search Team} Collaboration, A.~G. Riess et~al., {\it
  {Observational evidence from supernovae for an accelerating universe and a
  cosmological constant}},  {\em Astron. J.} {\bf 116} (1998) 1009--1038,
  [\href{http://arxiv.org/abs/astro-ph/9805201}{{\tt astro-ph/9805201}}].

\bibitem{Eisenstein_2005}
D.~J. Eisenstein, I.~Zehavi, D.~W. Hogg, et~al., {\it Detection of the baryon
  acoustic peak in the large‐scale correlation function of sdss luminous red
  galaxies},  {\em The Astrophysical Journal} {\bf 633} (Nov, 2005) 560–574.

\bibitem{Holz_2005}
D.~E. Holz and S.~A. Hughes, {\it Using gravitational‐wave standard sirens},
  {\em The Astrophysical Journal} {\bf 629} (Aug, 2005) 15–22.

\bibitem{Abbott_2021}
B.~P. Abbott, R.~Abbott, T.~D. Abbott, et~al., {\it A gravitational-wave
  measurement of the hubble constant following the second observing run of
  advanced ligo and virgo},  {\em The Astrophysical Journal} {\bf 909} (Mar,
  2021) 218.

\bibitem{Slosar:2019flp}
A.~Slosar et~al., {\it {Dark Energy and Modified Gravity}},
  \href{http://arxiv.org/abs/1903.12016}{{\tt arXiv:1903.12016}}.

\bibitem{Joyce:2014kja}
A.~Joyce, B.~Jain, J.~Khoury, and M.~Trodden, {\it {Beyond the Cosmological
  Standard Model}},  {\em Phys. Rept.} {\bf 568} (2015) 1--98,
  [\href{http://arxiv.org/abs/1407.0059}{{\tt arXiv:1407.0059}}].

\bibitem{Zlatev:1998tr}
I.~Zlatev, L.-M. Wang, and P.~J. Steinhardt, {\it {Quintessence, cosmic
  coincidence, and the cosmological constant}},  {\em Phys. Rev. Lett.} {\bf
  82} (1999) 896--899, [\href{http://arxiv.org/abs/astro-ph/9807002}{{\tt
  astro-ph/9807002}}].

\bibitem{Copeland:2006wr}
E.~J. Copeland, M.~Sami, and S.~Tsujikawa, {\it {Dynamics of dark energy}},
  {\em Int. J. Mod. Phys. D} {\bf 15} (2006) 1753--1936,
  [\href{http://arxiv.org/abs/hep-th/0603057}{{\tt hep-th/0603057}}].

\bibitem{Khoury:2003rn}
J.~Khoury and A.~Weltman, {\it {Chameleon cosmology}},  {\em Phys. Rev. D} {\bf
  69} (2004) 044026, [\href{http://arxiv.org/abs/astro-ph/0309411}{{\tt
  astro-ph/0309411}}].

\bibitem{Wagner:2012ui}
T.~Wagner, S.~Schlamminger, J.~Gundlach, and E.~Adelberger, {\it
  {Torsion-balance tests of the weak equivalence principle}},  {\em Class.
  Quant. Grav.} {\bf 29} (2012) 184002,
  [\href{http://arxiv.org/abs/1207.2442}{{\tt arXiv:1207.2442}}].

\bibitem{Adelberger:2003zx}
E.~G. Adelberger, B.~R. Heckel, and A.~E. Nelson, {\it {Tests of the
  gravitational inverse square law}},  {\em Ann. Rev. Nucl. Part. Sci.} {\bf
  53} (2003) 77--121, [\href{http://arxiv.org/abs/hep-ph/0307284}{{\tt
  hep-ph/0307284}}].

\bibitem{Burrage:2017qrf}
C.~Burrage and J.~Sakstein, {\it {Tests of Chameleon Gravity}},  {\em Living
  Rev. Rel.} {\bf 21} (2018), no.~1 1,
  [\href{http://arxiv.org/abs/1709.09071}{{\tt arXiv:1709.09071}}].

\bibitem{Noller:2020afd}
J.~Noller, {\it {Cosmological constraints on dark energy in light of
  gravitational wave bounds}},  {\em Phys. Rev. D} {\bf 101} (2020), no.~6
  063524, [\href{http://arxiv.org/abs/2001.05469}{{\tt arXiv:2001.05469}}].

\bibitem{Wang_2012}
J.~Wang, L.~Hui, and J.~Khoury, {\it No-go theorems for generalized chameleon
  field theories},  {\em Physical Review Letters} {\bf 109} (Dec, 2012).

\bibitem{Hinterbichler:2010es}
K.~Hinterbichler and J.~Khoury, {\it {Symmetron Fields: Screening Long-Range
  Forces Through Local Symmetry Restoration}},  {\em Phys. Rev. Lett.} {\bf
  104} (2010) 231301, [\href{http://arxiv.org/abs/1001.4525}{{\tt
  arXiv:1001.4525}}].

\bibitem{Vainshtein:1972sx}
A.~I. Vainshtein, {\it {To the problem of nonvanishing gravitation mass}},
  {\em Phys. Lett. B} {\bf 39} (1972) 393--394.

\bibitem{Nicolis:2008in}
A.~Nicolis, R.~Rattazzi, and E.~Trincherini, {\it {The Galileon as a local
  modification of gravity}},  {\em Phys. Rev. D} {\bf 79} (2009) 064036,
  [\href{http://arxiv.org/abs/0811.2197}{{\tt arXiv:0811.2197}}].

\bibitem{Babichev:2009ee}
E.~Babichev, C.~Deffayet, and R.~Ziour, {\it {k-Mouflage gravity}},  {\em Int.
  J. Mod. Phys. D} {\bf 18} (2009) 2147--2154,
  [\href{http://arxiv.org/abs/0905.2943}{{\tt arXiv:0905.2943}}].

\bibitem{Deffayet:2011gz}
C.~Deffayet, X.~Gao, D.~A. Steer, and G.~Zahariade, {\it {From k-essence to
  generalised Galileons}},  {\em Phys. Rev. D} {\bf 84} (2011) 064039,
  [\href{http://arxiv.org/abs/1103.3260}{{\tt arXiv:1103.3260}}].

\bibitem{Burrage_2015}
C.~Burrage, E.~J. Copeland, and J.~A. Stevenson, {\it Ellipticity weakens
  chameleon screening},  {\em Physical Review D} {\bf 91} (Mar, 2015).

\bibitem{Mota:2006fz}
D.~F. Mota and D.~J. Shaw, {\it {Evading Equivalence Principle Violations,
  Cosmological and other Experimental Constraints in Scalar Field Theories with
  a Strong Coupling to Matter}},  {\em Phys. Rev. D} {\bf 75} (2007) 063501,
  [\href{http://arxiv.org/abs/hep-ph/0608078}{{\tt hep-ph/0608078}}].

\bibitem{kelley1995iterative}
C.~T. Kelley, {\em Iterative methods for linear and nonlinear equations}.
\newblock SIAM, 1995.

\bibitem{AlnaesBlechta2015a}
M.~S. Aln{\ae}s, J.~Blechta, J.~Hake, et~al., {\it The fenics project version
  1.5},  {\em Archive of Numerical Software} {\bf 3} (2015), no.~100.

\bibitem{LoggMardalEtAl2012a}
A.~Logg, K.-A. Mardal, G.~N. Wells, et~al., {\em Automated Solution of
  Differential Equations by the Finite Element Method}.
\newblock Springer, 2012.

\bibitem{LoggWells2010a}
A.~Logg and G.~N. Wells, {\it Dolfin: Automated finite element computing},
  {\em ACM Transactions on Mathematical Software} {\bf 37} (2010), no.~2.

\bibitem{LoggWellsEtAl2012a}
A.~Logg, G.~N. Wells, and J.~Hake, {\em DOLFIN: a C++/Python Finite Element
  Library}, ch.~10.
\newblock Springer, 2012.

\bibitem{gmsh}
{Geuzaine, Christophe and Remacle, Jean-Francois}, ``Gmsh.''
  \href{http://http://gmsh.info/}{http://http://gmsh.info/}.

\bibitem{Geuzaine2009Gmsh}
C.~Geuzaine and J.-F. Remacle, {\it Gmsh: a three-dimensional finite element
  mesh generator with built-in pre- and post-processing facilities},  {\em
  International Journal for Numerical Methods in Engineering} {\bf 79} (2009),
  no.~11 1309--1331.

\bibitem{Braden_2021}
J.~Braden, C.~Burrage, B.~Elder, and D.~Saadeh, {\it $\varphi$enics: Vainshtein
  screening with the finite element method},  {\em Journal of Cosmology and
  Astroparticle Physics} {\bf 2021} (Mar, 2021) 010.

\bibitem{Burrage:2020bxp}
C.~Burrage, B.~Coltman, A.~Padilla, D.~Saadeh, and T.~Wilson, {\it {Massive
  Galileons and Vainshtein Screening}},  {\em JCAP} {\bf 02} (2021) 050,
  [\href{http://arxiv.org/abs/2008.01456}{{\tt arXiv:2008.01456}}].

\bibitem{Upadhye:2006vi}
A.~Upadhye, S.~S. Gubser, and J.~Khoury, {\it {Unveiling chameleons in tests of
  gravitational inverse-square law}},  {\em Phys. Rev. D} {\bf 74} (2006)
  104024, [\href{http://arxiv.org/abs/hep-ph/0608186}{{\tt hep-ph/0608186}}].

\bibitem{Elder:2016yxm}
B.~Elder, J.~Khoury, P.~Haslinger, et~al., {\it {Chameleon Dark Energy and Atom
  Interferometry}},  {\em Phys. Rev. D} {\bf 94} (2016), no.~4 044051,
  [\href{http://arxiv.org/abs/1603.06587}{{\tt arXiv:1603.06587}}].

\bibitem{Jaffe:2016fsh}
M.~Jaffe, P.~Haslinger, V.~Xu, et~al., {\it {Testing sub-gravitational forces
  on atoms from a miniature, in-vacuum source mass}},  {\em Nature Phys.} {\bf
  13} (2017) 938, [\href{http://arxiv.org/abs/1612.05171}{{\tt
  arXiv:1612.05171}}].

\bibitem{Elder:2019yyp}
B.~Elder, V.~Vardanyan, Y.~Akrami, et~al., {\it {Classical symmetron force in
  Casimir experiments}},  {\em Phys. Rev. D} {\bf 101} (2020), no.~6 064065,
  [\href{http://arxiv.org/abs/1912.10015}{{\tt arXiv:1912.10015}}].

\bibitem{Pernot-Borras:2020jev}
M.~Pernot-Borràs, J.~Bergé, P.~Brax, and J.-P. Uzan, {\it {Fifth force
  induced by a chameleon field on nested cylinders}},  {\em Phys. Rev. D} {\bf
  101} (2020), no.~12 124056, [\href{http://arxiv.org/abs/2004.08403}{{\tt
  arXiv:2004.08403}}].

\bibitem{Sabulsky_2019}
D.~O. Sabulsky, I.~Dutta, E.~A. Hinds, et~al., {\it Experiment to detect dark
  energy forces using atom interferometry},  {\em Physical Review Letters} {\bf
  123} (Aug, 2019).

\bibitem{langtangen2019introduction}
H.~Langtangen and K.~Mardal, {\em Introduction to Numerical Methods for
  Variational Problems}.
\newblock Texts in Computational Science and Engineering. Springer
  International Publishing, 2019.

\bibitem{LangtangenLogg2017}
H.~P. Langtangen and A.~Logg, {\em Solving PDEs in Python}.
\newblock Springer, 2017.

\bibitem{strang1973}
G.~Strang, {\it Piecewise polynomials and the finite element method},  {\em
  Bull. Amer. Math. Soc.} {\bf 79} (11, 1973) 1128--1137.

\bibitem{Burrage:2014oza}
C.~Burrage, E.~J. Copeland, and E.~Hinds, {\it {Probing Dark Energy with Atom
  Interferometry}},  {\em JCAP} {\bf 03} (2015) 042,
  [\href{http://arxiv.org/abs/1408.1409}{{\tt arXiv:1408.1409}}].

\bibitem{Ettori_2013}
S.~{Ettori}, A.~{Donnarumma}, E.~{Pointecouteau}, et~al., {\it {Mass Profiles
  of Galaxy Clusters from X-ray Analysis}},  {\em Space Science Reviews} {\bf
  177} (Aug., 2013) 119--154, [\href{http://arxiv.org/abs/1303.3530}{{\tt
  arXiv:1303.3530}}].

\bibitem{Applegate_2016}
D.~E. Applegate, A.~Mantz, S.~W. Allen, et~al., {\it {Cosmology and
  astrophysics from relaxed galaxy clusters – IV. Robustly calibrating
  hydrostatic masses with weak lensing}},  {\em Monthly Notices of the Royal
  Astronomical Society} {\bf 457} (02, 2016) 1522--1534,
  [\href{http://arxiv.org/abs/https://academic.oup.com/mnras/article-pdf/457/2/1522/2894538/stw005.pdf}{{\tt
  https://academic.oup.com/mnras/article-pdf/457/2/1522/2894538/stw005.pdf}}].

\bibitem{Lyapin_2019}
A.~R. {Lyapin} and R.~A. {Burenin}, {\it {Relation between X-ray and
  Sunyaev{\textemdash}Zeldovich Galaxy Cluster Mass Measurements}},  {\em
  Astronomy Letters} {\bf 45} (July, 2019) 403--410.

\bibitem{Terukina2014}
A.~{Terukina}, L.~{Lombriser}, K.~{Yamamoto}, et~al., {\it {Testing chameleon
  gravity with the Coma cluster}},  {\em JCAP} {\bf 2014} (Apr., 2014) 013,
  [\href{http://arxiv.org/abs/1312.5083}{{\tt arXiv:1312.5083}}].

\bibitem{Wilcox2015}
H.~{Wilcox}, D.~{Bacon}, R.~C. {Nichol}, et~al., {\it {The XMM Cluster Survey:
  testing chameleon gravity using the profiles of clusters}},  {\em MNRAS} {\bf
  452} (Sept., 2015) 1171--1183, [\href{http://arxiv.org/abs/1504.03937}{{\tt
  arXiv:1504.03937}}].

\bibitem{Sakstein2016}
J.~{Sakstein}, H.~{Wilcox}, D.~{Bacon}, K.~{Koyama}, and R.~C. {Nichol}, {\it
  {Testing gravity using galaxy clusters: new constraints on beyond Horndeski
  theories}},  {\em JCAP} {\bf 2016} (July, 2016) 019,
  [\href{http://arxiv.org/abs/1603.06368}{{\tt arXiv:1603.06368}}].

\bibitem{Koyama2016}
K.~{Koyama}, {\it {Cosmological tests of modified gravity}},  {\em Reports on
  Progress in Physics} {\bf 79} (Apr., 2016) 046902,
  [\href{http://arxiv.org/abs/1504.04623}{{\tt arXiv:1504.04623}}].

\bibitem{Burrage2018}
C.~{Burrage} and J.~{Sakstein}, {\it {Tests of chameleon gravity}},  {\em
  Living Reviews in Relativity} {\bf 21} (Mar., 2018) 1,
  [\href{http://arxiv.org/abs/1709.09071}{{\tt arXiv:1709.09071}}].

\bibitem{Cataneo2018}
M.~{Cataneo} and D.~{Rapetti}, {\it {Tests of gravity with galaxy clusters}},
  {\em International Journal of Modern Physics D} {\bf 27} (Jan., 2018)
  1848006--936, [\href{http://arxiv.org/abs/1902.10124}{{\tt
  arXiv:1902.10124}}].

\bibitem{Tamosiunas2019}
A.~{Tamosiunas}, D.~{Bacon}, K.~{Koyama}, and R.~C. {Nichol}, {\it {Testing
  emergent gravity on galaxy cluster scales}},  {\em JCAP} {\bf 2019} (May,
  2019) 053, [\href{http://arxiv.org/abs/1901.05505}{{\tt arXiv:1901.05505}}].

\bibitem{tamosiunas2021chameleon}
A.~Tamosiunas, C.~Briddon, C.~Burrage, W.~Cui, and A.~Moss, {\it Chameleon
  screening depends on the shape and structure of {NFW} halos},
  \href{http://arxiv.org/abs/2108.10364}{{\tt arXiv:2108.10364}}.

\bibitem{Navarro1997}
J.~F. {Navarro}, C.~S. {Frenk}, and S.~D.~M. {White}, {\it {A Universal Density
  Profile from Hierarchical Clustering}},  {\em ApJ} {\bf 490} (Dec., 1997)
  493--508, [\href{http://arxiv.org/abs/astro-ph/astro-ph/9611107}{{\tt
  arXiv:astro-ph/astro-ph/9611107}}].

\bibitem{Burrage:2017shh}
C.~Burrage, E.~J. Copeland, A.~Moss, and J.~A. Stevenson, {\it {The shape
  dependence of chameleon screening}},  {\em JCAP} {\bf 01} (2018) 056,
  [\href{http://arxiv.org/abs/1711.02065}{{\tt arXiv:1711.02065}}].

\bibitem{Hamilton:2015zga}
P.~Hamilton, M.~Jaffe, P.~Haslinger, et~al., {\it {Atom-interferometry
  constraints on dark energy}},  {\em Science} {\bf 349} (2015) 849--851,
  [\href{http://arxiv.org/abs/1502.03888}{{\tt arXiv:1502.03888}}].

\bibitem{Sabulsky:2018jma}
D.~O. Sabulsky, I.~Dutta, E.~A. Hinds, et~al., {\it {Experiment to detect dark
  energy forces using atom interferometry}},  {\em Phys. Rev. Lett.} {\bf 123}
  (2019), no.~6 061102, [\href{http://arxiv.org/abs/1812.08244}{{\tt
  arXiv:1812.08244}}].

\bibitem{Brax:2014gja}
P.~Brax, {\it {Testing Chameleon Fields with Ultra Cold Neutron Bound States
  and Neutron Interferometry}},  {\em Phys. Procedia} {\bf 51} (2014) 73--77.

\bibitem{Jenke:2014yel}
T.~Jenke et~al., {\it {Gravity Resonance Spectroscopy Constrains Dark Energy
  and Dark Matter Scenarios}},  {\em Phys. Rev. Lett.} {\bf 112} (2014) 151105,
  [\href{http://arxiv.org/abs/1404.4099}{{\tt arXiv:1404.4099}}].

\bibitem{Lemmel:2015kwa}
H.~Lemmel, P.~Brax, A.~N. Ivanov, et~al., {\it {Neutron Interferometry
  constrains dark energy chameleon fields}},  {\em Phys. Lett. B} {\bf 743}
  (2015) 310--314, [\href{http://arxiv.org/abs/1502.06023}{{\tt
  arXiv:1502.06023}}].

\bibitem{Li:2016tux}
K.~Li et~al., {\it {Neutron Limit on the Strongly-Coupled Chameleon Field}},
  {\em Phys. Rev. D} {\bf 93} (2016), no.~6 062001,
  [\href{http://arxiv.org/abs/1601.06897}{{\tt arXiv:1601.06897}}].

\bibitem{Sponar:2021rvk}
S.~Sponar, R.~I.~P. Sedmik, M.~Pitschmann, H.~Abele, and Y.~Hasegawa, {\it
  {Tests of fundamental quantum mechanics and dark interactions with low-energy
  neutrons}},  {\em Nature Rev. Phys.} {\bf 3} (2021), no.~5 309--327.

\bibitem{Jenke:2020obe}
T.~Jenke, J.~Bosina, J.~Micko, et~al., {\it {Gravity resonance spectroscopy and
  dark energy symmetron fields: qBOUNCE experiments performed with Rabi and
  Ramsey spectroscopy}},  {\em Eur. Phys. J. ST} {\bf 230} (2021), no.~4
  1131--1136, [\href{http://arxiv.org/abs/2012.07472}{{\tt arXiv:2012.07472}}].

\bibitem{Upadhye:2012qu}
A.~Upadhye, {\it {Dark energy fifth forces in torsion pendulum experiments}},
  {\em Phys. Rev. D} {\bf 86} (2012) 102003,
  [\href{http://arxiv.org/abs/1209.0211}{{\tt arXiv:1209.0211}}].

\bibitem{Pernot-Borras:2021edr}
M.~Pernot-Borr\`as, J.~Berg\'e, P.~Brax, et~al., {\it {Constraints on chameleon
  gravity from the measurement of the electrostatic stiffness of the MICROSCOPE
  mission accelerometers}},  {\em Phys. Rev. D} {\bf 103} (2021), no.~6 064070,
  [\href{http://arxiv.org/abs/2102.00023}{{\tt arXiv:2102.00023}}].

\bibitem{Zhao:2021anp}
Y.-L. Zhao, Y.-J. Tan, W.-H. Wu, J.~Luo, and C.-G. Shao, {\it {Constraining the
  chameleon model with the HUST-2020 torsion pendulum experiment}},  {\em Phys.
  Rev. D} {\bf 103} (2021), no.~10 104005.

\bibitem{Decca:2007jq}
R.~S. Decca, D.~Lopez, E.~Fischbach, et~al., {\it {Novel constraints on light
  elementary particles and extra-dimensional physics from the Casimir effect}},
   {\em Eur. Phys. J. C} {\bf 51} (2007) 963--975,
  [\href{http://arxiv.org/abs/0706.3283}{{\tt arXiv:0706.3283}}].

\bibitem{Almasi:2015zpa}
A.~Almasi, P.~Brax, D.~Iannuzzi, and R.~I.~P. Sedmik, {\it {Force sensor for
  chameleon and Casimir force experiments with parallel-plate configuration}},
  {\em Phys. Rev. D} {\bf 91} (2015), no.~10 102002,
  [\href{http://arxiv.org/abs/1505.01763}{{\tt arXiv:1505.01763}}].

\bibitem{Sedmik:2018kqt}
R.~Sedmik and P.~Brax, {\it {Status Report and first Light from Cannex: Casimir
  Force Measurements between flat parallel Plates}},  {\em J. Phys. Conf. Ser.}
  {\bf 1138} (2018), no.~1 012014.

\bibitem{Klimchitskaya:2021lak}
G.~L. Klimchitskaya and V.~M. Mostepanenko, {\it {Dark Matter Axions,
  Non-Newtonian Gravity and Constraints on them from Recent Measurement of the
  Casimir Force in the Micrometer Separation Range}},  {\em Universe} {\bf 7}
  (9, 2021) N9, [\href{http://arxiv.org/abs/2109.06534}{{\tt
  arXiv:2109.06534}}].

\bibitem{Sedmik:2021iaw}
R.~I.~P. Sedmik and M.~Pitschmann, {\it {Next Generation Design and Prospects
  for Cannex}},  {\em Universe} {\bf 7} (2021), no.~7 234,
  [\href{http://arxiv.org/abs/2107.07645}{{\tt arXiv:2107.07645}}].

\bibitem{Geraci:2010ft}
A.~A. Geraci, S.~B. Papp, and J.~Kitching, {\it {Short-range force detection
  using optically-cooled levitated microspheres}},  {\em Phys. Rev. Lett.} {\bf
  105} (2010) 101101, [\href{http://arxiv.org/abs/1006.0261}{{\tt
  arXiv:1006.0261}}].

\bibitem{Rider:2016xaq}
A.~D. Rider, D.~C. Moore, C.~P. Blakemore, et~al., {\it {Search for Screened
  Interactions Associated with Dark Energy Below the 100 $\mathrm{\mu m}$
  Length Scale}},  {\em Phys. Rev. Lett.} {\bf 117} (2016), no.~10 101101,
  [\href{http://arxiv.org/abs/1604.04908}{{\tt arXiv:1604.04908}}].

\bibitem{Liu:2018jia}
J.~Liu and K.-D. Zhu, {\it {Cavity optomechanical spectroscopy constraints
  chameleon dark energy scenarios}},  {\em Eur. Phys. J. C} {\bf 78} (2018),
  no.~3 266.

\bibitem{Qvarfort:2021zrl}
S.~Qvarfort, D.~R\"atzel, and S.~Stopyra, {\it {Constraining modified gravity
  with quantum optomechanics}},  \href{http://arxiv.org/abs/2108.00742}{{\tt
  arXiv:2108.00742}}.

\bibitem{Silvestri:2011ch}
A.~Silvestri, {\it {Scalar radiation from Chameleon-shielded regions}},  {\em
  Phys. Rev. Lett.} {\bf 106} (2011) 251101,
  [\href{http://arxiv.org/abs/1103.4013}{{\tt arXiv:1103.4013}}].

\bibitem{Sakstein:2013pda}
J.~Sakstein, {\it {Stellar Oscillations in Modified Gravity}},  {\em Phys. Rev.
  D} {\bf 88} (2013), no.~12 124013,
  [\href{http://arxiv.org/abs/1309.0495}{{\tt arXiv:1309.0495}}].

\bibitem{Upadhye:2013nfa}
A.~Upadhye and J.~H. Steffen, {\it {Monopole radiation in modified gravity}},
  \href{http://arxiv.org/abs/1306.6113}{{\tt arXiv:1306.6113}}.

\bibitem{Brax:2013uh}
P.~Brax, A.-C. Davis, and J.~Sakstein, {\it {Pulsar Constraints on Screened
  Modified Gravity}},  {\em Class. Quant. Grav.} {\bf 31} (2014) 225001,
  [\href{http://arxiv.org/abs/1301.5587}{{\tt arXiv:1301.5587}}].

\bibitem{Hagala:2016fks}
R.~Hagala, C.~Llinares, and D.~F. Mota, {\it {Cosmic Tsunamis in Modified
  Gravity: Disruption of Screening Mechanisms from Scalar Waves}},  {\em Phys.
  Rev. Lett.} {\bf 118} (2017), no.~10 101301,
  [\href{http://arxiv.org/abs/1607.02600}{{\tt arXiv:1607.02600}}].

\bibitem{Ikeda:2021pkb}
T.~Ikeda, V.~Cardoso, and M.~Zilh\~ao, {\it {Instabilities of scalar fields
  around oscillating stars}},  \href{http://arxiv.org/abs/2110.06937}{{\tt
  arXiv:2110.06937}}.

\end{thebibliography}\endgroup

\end{document}